\documentclass[12pt]{article}
\usepackage{amsmath,amssymb,amsfonts,amsthm,bbm}
\usepackage{epic,eepic,epsfig,longtable}
\usepackage{multirow,verbatim}
\usepackage{array}
\usepackage{graphicx}
\usepackage{floatrow}
\usepackage{subcaption}
\usepackage{paralist}
\usepackage{latexsym}
\usepackage{comment}
\usepackage{epsfig}
\usepackage{setspace}
\usepackage{CJK}
\usepackage{color}

\usepackage{geometry}
\usepackage{algorithm}
\usepackage{algpseudocode}

\usepackage{rotating}

\usepackage[authoryear,round]{natbib}
\def\spacingset#1{\renewcommand{\baselinestretch}%
{#1}\small\normalsize} \spacingset{1}

\makeatletter
\def\singlespace{\def\baselinestretch{1}\@normalsize}

\numberwithin{equation}{section}

\renewcommand{\hat}{\widehat}

\renewcommand{\hat}{\widehat}

\newcommand{\bfm}[1]{\ensuremath{\mathbf{#1}}}

   \def\bA{\bfm A}  
   \def\bB{\bfm B}

    \def\FF{\mathbb{F}}

   \def\bI{\bfm I}  
   \def\bJ{\bfm J}

\def\bw{\bfm w}   \def\bW{\bfm W}  
\def\bx{\bfm x}   \def\bX{\bfm X}

\newcommand{\bfsym}[1]{\ensuremath{\boldsymbol{#1}}}

 \def\bbeta{\bfsym \beta}

 \def\bmu{\bfsym {\mu}}                 
 \def\bnu{\bfsym {\nu}}

 \def\bsigma{\bfsym \sigma}             \def\bSigma{\bfsym \Sigma}
         \def\bLambda {\bfsym {\Lambda}}
           \def\bOmega {\bfsym {\Omega}}

\def\1{\bfsym{1}}	

\DeclareMathOperator{\argmin}{argmin}

\DeclareMathOperator{\tr}{tr}

\def\newpage{\vfill\eject}

\def\today{\ifcase\month\or
  January\or February\or March\or April\or May\or June\or
  July\or August\or September\or October\or November\or December\fi
  \space\number\day, \number\year}

\newdimen\biblioindent    \biblioindent=30pt

 at 8truept

\newcommand{\beq}{\begin{equation}}
  \newcommand{\eeq}{\end{equation}}
\newcommand{\beqn}{\begin{eqnarray}}
  \newcommand{\eeqn}{\end{eqnarray}}
\newcommand{\beqnn}{\begin{eqnarray*}}
  \newcommand{\eeqnn}{\end{eqnarray*}}

\def\lasso{{\rm LASSO}}
\def\svr{{\rm SVR}}

\allowdisplaybreaks
\usepackage{verbatim}
\def\tilde{\widetilde}

\def\FF{\mathcal{F}}
\def\[{\left [}  \def\]{\right ]} \def\({\left (}  \def\){\right )}
 \def\endpf{$\blacksquare$}
\def\hat{\widehat}

\newtheorem{assumption}{Assumption}
\newtheorem{theorem}{Theorem}

\newtheorem{proposition}{Proposition}
\theoremstyle{definition}

\newtheorem{remark}{Remark}

\title{Robust High-Dimensional Time-Varying Coefficient Estimation}
\author{Minseok Shin$^a$ and Donggyu Kim$^b$\footnote{Corresponding author. E-mail address: donggyu.kim@ucr.edu.} \\
$^a$Pohang University of Science and Technology (POSTECH) \\
$^b$University of California, Riverside\\
}

\begin{document}

\maketitle
\begin{spacing}{1.9}

\begin{abstract}
In this paper, we develop a novel high-dimensional coefficient estimation procedure based on high-frequency data.
Unlike usual high-dimensional regression procedures such as LASSO, we additionally handle the heavy-tailedness of high-frequency observations as well as time variations of coefficient processes.
Specifically, we employ the Huber loss and a truncation scheme to handle heavy-tailed observations, while $\ell_{1}$-regularization is adopted to overcome the curse of dimensionality.
To account for the time-varying coefficient, we estimate local coefficients which are biased due to the  $\ell_{1}$-regularization. 
Thus, when estimating integrated coefficients, we propose a debiasing scheme to enjoy the law of large numbers property and employ a thresholding scheme to further accommodate the sparsity of the coefficients. 
We call this Robust thrEsholding Debiased LASSO (RED-LASSO) estimator.
We show that the RED-LASSO estimator can achieve a near-optimal convergence rate.
In the empirical study, we apply the RED-LASSO procedure to the high-dimensional integrated coefficient estimation using high-frequency trading data.
\end{abstract}


\noindent \textbf{Keywords:} Debias,  diffusion process, LASSO, factor model, sparsity, Huber loss, heavy-tail.

\section{Introduction} \label{SEC-1}
With the wide availability of high-frequency financial data, researchers have developed financial models that can incorporate high-frequency data, and empirical studies have shown that these models better account for market dynamics.  
For example, auto-regressive-type models  have been introduced based on high-frequency-based measures, such as  realized volatility  and realized beta estimators \citep{andersen2006realized, corsi2009simple, engle2006multiple, hansen2012realized, kim2016unified, kim2019factor, shephard2010realising, song2021volatility}.
Empirical studies have demonstrated that capturing the auto-regressive structures of high-frequency measures helps explain financial market dynamics.
On the other hand, we often employ realized volatility estimators when analyzing regression models, such as the Capital Asset Pricing Model (CAPM) \citep{lintner1965security, sharpe1964capital} and multi-factor models \citep{fama1992cross}.
For example, market beta can be estimated by a ratio of the realized covariance between assets and systematic factors to the realized variance of the systematic factors \citep{barndorff2004econometric}. 
See \citet{andersen2006realized, mykland2009inference, reiss2015nonparametric} for the related literature.
\citet{li2017adaptive} derived the asymptotic efficiency bound for betas in a linear continuous-time regression model.  
In addition, empirical studies have shown the time-varying feature of the beta process \citep{ang2012testing, ferson1999conditioning, kalnina2022inference, kong2023discrepancy, kong2018testing, oh2024robust,  reiss2015nonparametric}.
To address this issue, \citet{ait2020high} employed time-localized regressions for the multi-factor models.
\citet{chen2018inference} introduced the general nonparametric inference for nonlinear volatility functionals of general multivariate It\^o semimartingales.
These models and estimation methods have shown that incorporating high-frequency data helps better account for the beta dynamics in the finite-dimensional set-up.

In modern financial studies and practices, researchers have found a large number of factor candidates \citep{bali2011maxing, campbell2008search, cochrane2011presidential, harvey2016and, hou2020replicating, mclean2016does}. 
Thus, we often encounter the curse of dimensionality,  and the beta estimation methods designed for the finite dimension are neither efficient nor effective.
To handle the high-dimensionality, we often employ  LASSO \citep{tibshirani1996regression}, SCAD \citep{fan2001variable}, and the Dantzig selector \citep{candes2007dantzig} under the sparsity condition of model parameters.
However, direct application of these methods cannot handle the time-varying feature of beta processes.
Recently,  \citet{Kim2024regression} developed a  Thresholded dEbiased Dantzig (TED) estimator  that can handle the high-dimensionality and time variation of beta processes.   
Specifically, they employed the Dantzig selector \citep{candes2007dantzig} for each time window and estimated the integrated beta with the debiasing and truncation schemes.
They established the asymptotic properties of the TED estimator under the sub-Gaussianity assumption on the high-frequency log-return data.
However, we often observe that high-frequency financial data exhibit heavy tails \citep{cont2001empirical, fan2018robust, mao2018stochastic, shin2023adaptive}.
Under the heavy-tailedness assumption, the existing estimation methods, including the TED estimator \citep{Kim2024regression},  cannot consistently  estimate the time-varying betas.
Specifically, they fail to control the tail behavior of the local beta estimator and the bias adjustment term, which can lead to large estimation errors.
These facts lead to the demand for developing methodologies that can simultaneously handle heavy-tailed observations, the curse of dimensionality, and time-varying beta processes.

In this paper, we develop a robust integrated beta estimator based on high-dimensional regression jump-diffusion processes.
To handle the high-dimensionality and time-varying beta, we assume that the beta processes are sparse and follow a continuous diffusion process.
To account for the heavy-tailedness of financial data, we assume that the residual process and jump size processes satisfy only a finite $\(2+\zeta\)$-th moment condition for an arbitrarily small $\zeta>0$.
That is, we assume that the sources of the heavy-tailedness are the residual process and jump. 
We first estimate the instantaneous betas as follows.
We employ the $\ell_1$-penalty, Huber loss, and truncation method to manage the curse of dimensionality, heavy-tailedness of the residual process, and jumps, respectively. 
We show that the proposed instantaneous beta estimator has the desirable convergence rate.
However, the instantaneous beta estimator has non-negligible biases coming from the Huber loss and $\ell_1$-penalty.
Thus, to estimate the integrated beta using the instantaneous beta estimators, we need to mitigate the biases. 
Since the biases are heavy-tailed, the existing debiasing scheme cannot efficiently adjust the biases.
To tackle this problem, we propose a novel debiasing scheme and obtain an integrated beta estimator.
We show that the debiased integrated beta estimator has a near-optimal convergence rate and outperforms the simple integration of the instantaneous beta estimators without a debiasing scheme. 
However, due to the bias adjustment, the debiased integrated beta estimator is not sparse; thus, we further regularize it to accommodate the sparsity. 
We call this the Robust thrEsholding Debiased LASSO (RED-LASSO)  estimator.
We also show that the RED-LASSO estimator has a near-optimal convergence rate.

The rest of the paper is organized as follows.
Section \ref{SEC-2}  introduces the high-dimensional regression jump-diffusion process.
Section \ref{SEC-3} proposes the RED-LASSO estimator and establishes its asymptotic properties.
In Section \ref{SEC-4}, we conduct a simulation study to check the finite sample performance of the proposed estimation method. 
In Section \ref{SEC-5}, we apply the proposed estimation procedure to high-frequency financial data.
The conclusion is presented in Section \ref{SEC-6}, and all of the proofs are collected in the Appendix.

\setcounter{equation}{0}
\section{The model set-up} \label{SEC-2}
We first fix some notations.
For any given  $p_1$ by $ p_2$ matrix $\bA = \left(A_{ij}\right)$, let
 \begin{equation*}
  	 \|\bA\|_1 = \max\limits_{1 \leq j \leq p_2}\sum\limits_{i = 1}^{p_1}|A_{ij}|, \quad \|\bA\|_\infty = \max\limits_{1 \leq i \leq p_1}\sum\limits_{j = 1}^{p_2}|A_{ij}|, \quad \text{ and } \quad \| \bA \| _{\max} = \max_{i,j} | A_{ij}|.
 \end{equation*}
The Frobenius norm of $\bA$ is denoted by $\|\bA\|_F = \sqrt{ \mathrm{tr}(\bA^{\top} \bA) }$ and the matrix spectral norm $\|\bA\|_2$ is  the square root of the largest eigenvalue of $\bA\bA^\top$.
We will use $C$'s to denote generic constants whose values are free of $n$ and $p$ and may change from appearance to appearance.

Let $Y(t)$ and $\bX(t)=\(X_1(t), \ldots, X_p(t)\)^{\top}$  be the dependent process and $p$-dimensional multivariate covariate process, respectively.
We employ the following non-parametric time-series regression jump-diffusion model:
\begin{eqnarray} \label{model-1}
&& dY(t)=   dY^c(t) + dY^J(t), \cr
&& dY^c(t)=\bbeta ^{\top}(t)  d\bX^c(t)+d Z^c(t), \quad \text{and} \quad dY^J(t) = J^y(t) d \Lambda^y(t),  
\end{eqnarray}
where $Y^c(t)$ and $\bX^c(t)=\(X^c_1(t), \ldots, X^c_p(t)\)^{\top}$ are the continuous parts of $Y(t)$ and $\bX(t)$, respectively, $Y^J(t)$ is the jump part of $Y(t)$,  $J^y(t)$ is a jump size, $\Lambda^y(t)$ is a Poisson process with a bounded intensity process, $\bbeta(t)=\(\beta_1(t), \ldots, \beta_p(t)\)^{\top}$ is a coefficient process, and $Z^c(t)$ is a residual process. 
We note that the subscript $c$ represents the continuous part of the process. 
The covariate process $\bX(t)$ and residual process $Z^c(t)$ satisfy
\begin{eqnarray}\label{model-2}
	&&d\bX(t)=  d \bX^c(t) + d \bX^J(t),   \quad  d\bX^c(t)= \bmu(t) dt+ \bsigma(t) d\bB(t), \cr
	&&  d \bX^J(t) =  \bJ(t) d \bLambda(t), \quad \text{and} \quad    dZ^c(t)= \nu(t) d W(t),
\end{eqnarray}
where $\bX^J(t)$ is the jump part of $\bX(t)$, $\bJ(t)=\(J_1(t), \ldots, J_p(t)\)^{\top}$ is a jump size process, $\bLambda(t)$ is a $p$-dimensional Poisson process with bounded intensity processes, $\bsigma(t)$ is a $p$ by $q$ matrix, and $\bB(t)$ and $W(t)$ are $q$-dimensional and one-dimensional independent Brownian motions, respectively.
The stochastic processes $\bmu(t)$, $\bbeta(t)$, $\bsigma(t)$, and $\nu(t)$ are defined on a filtered probability space $(\Omega, \FF, \{\FF_t, t \in [0, 1]\},  P)$ with filtration $\FF_t$ satisfying the usual conditions, such as adapted and c\`adl\`ag process.
In this paper, we do not assume that $\nu(t)$ is bounded.
Instead, we only impose the finite moment condition on the residual process in Assumption \ref{assumption1}(a), which allows the residual process to exhibit heavy tails.
We assume that the coefficient $\bbeta(t)=\left(\beta_{1}(t), \ldots, \beta_{p}(t)\right)^{\top}$ satisfies the following diffusion model:
\begin{equation*}
	d \bbeta(t)= \bmu_{\beta}(t) dt + \bnu_{\beta}(t) d \bW_{\beta}(t),
\end{equation*}
where $\bnu_{\beta}(t)$ is a $p$ by $r$ matrix, $\bW_{\beta}(t)$ is a $r$-dimensional independent Brownian motion, and $\bmu_{\beta}(t)$ and $\bnu_{\beta}(t)$ are predictable.
The main interest of this paper is to investigate the latent regression diffusion process. 
From this point of view, the jump part can be considered as noises, and we discuss how to overcome this in the following section. 
The parameter of interest is the integrated beta:
\begin{equation*}	
	I \beta = (I \beta_i)_{i=1,\ldots, p}= \int_{0}^1 \bbeta(t) dt. 
\end{equation*}
The integrated beta can be considered as the average of spot betas. 
That is, the integrated beta presents the average effect of the increment of the covariate process.
When the beta process is constant, the integrated beta is the same as the usual beta in the regression model. 
\begin{remark}
In this paper, we focus on $\int_{0}^1 \bbeta(t) dt$ rather than $\int_{0}^1 \left|\bbeta(t)\right| dt$.
Then, the variable selection is based on the average effect of the covariate process on the dependent process.
For example, suppose that the beta process smoothly oscillates around zero.
In this case, $\int_{0}^1 \bbeta(t)\,dt$ would be close to zero, whereas $\int_{0}^1 \left|\bbeta(t)\right| dt$ can be large.
Such factors with zero average effect are excluded in applications that emphasize the overall long-term effect.
We note that the drift term $\bmu_{\beta}(t)=\left(\mu_{\beta,1}(t), \ldots, \mu_{\beta,p}(t)\right)^{\top}$ can play a significant role in the difference between $\int_{0}^1 \bbeta(t) dt$ and $\int_{0}^1 \left|\bbeta(t)\right| dt$.
For example, suppose that $\beta_i(0)=0$.
When $\mu_{\beta,i}(t)$ changes sign over time, $\beta_i(t)$ may oscillate around zero, which can lead to a significant difference between the two measures.
In contrast, when $\mu_{\beta,i}(t)$ maintains the same sign and $\beta_i(t)$ fluctuates around that level due to the stochastic Brownian motion component, the difference is relatively small.
On the other hand, when the beta process exhibits discontinuities or nonsmoothness, $\int_{0}^1 \left|\bbeta(t)\right| dt$  can serve as a more relevant measure.
However, addressing such cases is beyond the scope of this paper, and theoretically, the localization scheme does not work.
Thus, we leave this issue for a future study.
\end{remark}

In the regression-based financial models, there are hundreds of potential factor candidates \citep{bali2011maxing, campbell2008search, cochrane2011presidential, harvey2016and, hou2020replicating, mclean2016does}. 
To account for this, we allow that the dimension $p$ can be large; thus, we need to handle the curse of dimensionality.
To do this, we assume that the coefficient beta process $\bbeta(t) = (\beta_{1}(t) ,\ldots, \beta_{p}(t)) ^{\top}$ satisfies the following sparsity condition:
\begin{equation}\label{sparsity_beta}
	\sup_{0 \leq t \leq 1}\sum_{i=1}^p    |\beta_{i}(t) | ^{\delta}  \leq s_p  \quad \text{and} \quad 	\sum_{i=1}^p|I  \beta_{i} | ^{\delta}  \leq s_p \, \text{ a.s.},
\end{equation}
where $ \delta \in [0,1)$, $s_p$ is diverging slowly in $p$, and $0^0$ is defined as 0. 
This general sparsity condition includes the exact sparsity condition, i.e., $\delta=0$.
The exact sparsity condition implies that only several factors are significant, while most factors do not affect the dependent process.
Thus, we assume that the relatively small number of factors is significant.  
We note that since the beta process is an It\^o diffusion process, in general, the boundedness in the sparsity condition \eqref{sparsity_beta} is satisfied with high probability.
Thus, even without the almost sure sparsity condition, the results in this paper hold with high probability.
However, for simplicity, we assume that the sparsity condition holds almost surely.
%

\section{Robust high-dimensional high-frequency regression}\label{SEC-3}
\subsection{Integrated beta estimation procedure} \label{SEC-Estimation}
In this section, we propose a robust integrated beta estimation procedure for the high-dimensional regression diffusion model defined in  \eqref{model-1}--\eqref{model-2}.
Recently, under the sub-Gaussian assumption, \citet{Kim2024regression} proposed the integrated beta estimator that can handle the curse of dimensionality and time-varying betas.
However,  empirical studies have demonstrated that the stock log-return data often exhibit heavy-tails \citep{cont2001empirical, fan2018robust, mao2018stochastic, shin2023adaptive}, which leads to the inconsistency of the integrated beta estimator.
To accommodate heavy-tailedness, we impose the finite moment condition on the residual process, $Z^c(t)$, and jump sizes, $J^y(t)$ and $\bJ(t)$ (see Assumption \ref{assumption1}).
Then, we propose a robust estimation procedure.
We first estimate the instantaneous betas.
To do this, we employ the local regression as follows.
For any process $g(t)$ and $\Delta_n=1/n$, let $\Delta_i ^n g = g(i \Delta_n) - g((i-1) \Delta_{n})$ for $1 \leq i \leq 1/\Delta_n$.
Define 
\begin{eqnarray*}
&& \mathcal{Y}_i = 
\begin{pmatrix}
\Delta_{i+1} ^n Y  \\ 
\Delta_{i+2} ^n Y  \\  
 \vdots \\ 
\Delta_{i+k_n} ^n Y 
\end{pmatrix},
\quad 
\mathcal{Z}_i = 
\begin{pmatrix}
\Delta_{i+1} ^n Z^c \\ 
\Delta_{i+2} ^n Z^c \\  
 \vdots \\ 
\Delta_{i+k_n} ^n Z^c
\end{pmatrix},
 \cr
\\ 
&&  \mathcal{X}_i = 
\begin{pmatrix}
\Delta_{i+1} ^n \hat{\bX}^{c\top} \\ 
\Delta_{i+2} ^n \hat{\bX}^{c\top}\\  
 \vdots \\ 
\Delta_{i+k_n} ^n \hat{\bX}^{c\top}
\end{pmatrix}, 
\quad \text{and} \quad
\Delta_{i} ^n \hat{\bX}^c =
\begin{pmatrix}
\Delta_{i} ^n X_1\, \1_{\{ | \Delta_{i} ^n X_1 | \leq v_{1,n} \}} \\ 
\Delta_{i} ^n X_2\, \1_{\{ | \Delta_{i} ^n X_2 | \leq v_{2,n} \}} \\  
 \vdots \\ 
\Delta_{i} ^n X_p\, \1_{\{ | \Delta_{i} ^n X_p | \leq v_{p,n} \}}
\end{pmatrix},
\end{eqnarray*}
where $k_n$ is the number of observations for each local regression, $\1_{\{  \cdot \} }$ is an indicator function, and $v_{j,n}$, $j=1, \ldots, p$, are the threshold levels. 
We use $v_{j,n} = C_{j,v}  \sqrt{\log p} n^{-1/2}$ for some large constants $C_{j,v}$, $j=1, \ldots, p$.
In the numerical study, we choose
\begin{equation}\label{jump_adj} 
  v_{j,n}= \sqrt{BV_j \log p} n^{-1/2},
\end{equation}
where the bipower variation $BV_j = \dfrac{\pi}{2}\sum_{i=2}^{n} | \Delta_{i-1} ^n X_j| \cdot | \Delta_{i} ^n X_j|$.
This choice of $v_{j,n}$ is similar to the usual choice in the literature \citep{ait2020high, ait2019principal} except for the $\log p$ term, which is used to bound the continuous parts of the covariate processes with high probability.
We note that the thresholding can detect the jumps in the covariate process $X(t)$ and mitigate their impact on beta estimators. 
On the other hand, the thresholding is not used for the dependent process $Y(t)$ since the robustification method outlined in \eqref{loss_func} and \eqref{debias}  can handle both heavy-tailedness of the residual process $Z^c(t)$ and jumps in the dependent process $Y(t)$. 
\begin{remark}
In this paper, we handle jumps using thresholding and robustification methods under the finite activity assumption.
When extending this assumption to allow for jumps of infinite activity or infinite variation, a major challenge arises from the presence of numerous small jumps.
To address this issue, we can apply truncation methods.
Specifically, using truncation techniques, we can establish a CLT for the estimated volatility functionals in the presence of infinite activity but finite variation jumps \citep{mancini2009non, mancini2017truncated}.
For the infinite variation case, we can still obtain consistency, although the convergence rate is determined by the jump activity indices \citep{mancini2017truncated}.
We note that these CLT and consistency results are established in the finite-dimensional setting.
However, it is a theoretically demanding task to extend these results to the high-dimensional case.
Thus, we leave this issue for future study.
\end{remark}

Meanwhile, when calculating local regressions, we need to handle the curse of dimensionality and heavy-tailedness.
To overcome high-dimensionality, we often employ the penalized regression procedures under the sparsity assumption.
 For example, we often use the LASSO \citep{tibshirani1996regression} and Dantzig \citep{candes2007dantzig} estimators with the sub-Gaussian conditions. 
 However, these estimators cannot handle the heavy-tailed observations, and furthermore, they are not consistent. 
 To tackle this issue, we use the following Huber loss $l_{\tau}$ \citep{huber1964robust}:
\begin{equation*}
l_{\tau}\left( x\right) =\begin{cases}x^{2}/2 & \text{ if } \left| x\right| \leq \tau \\ \tau\left| x\right| -\tau^{2}/2 & \text{ if } \left| x\right| > \tau, \end{cases}
\end{equation*}
where $\tau>0$ is the robustification parameter. 
We denote $l_{\tau}\left( \bx\right)=\left(l_{\tau}\left(x_{1}\right), \ldots, l_{\tau}\left( x_{p_1}\right)\right)^{\top}$ for any vector $\bx = \left(x_1, \ldots, x_{p_1} \right)^{\top} \in \mathbb{R}^{p_1}$.
The Huber loss $l_{\tau}$ mitigates the effect of outliers coming from the heavy-tailedness of the residual process $Z^c(t)$ and jump size process $J^y(t)$. 
Thus, by employing the truncation, Huber loss, and $\ell_{1}$-regularization, we can simultaneously deal with the three issues of the jumps, heavy-tailedness, and curse of dimensionality.
Specifically, we propose the following instantaneous beta estimator at time $i \Delta_n$:
\begin{equation}\label{ins_beta}
 \hat{\bbeta}_{i \Delta_n} = \arg \min_{\bbeta \in \mathbb{R}^{p}}{\mathcal{L}}_{\tau,i}(\bbeta) + \eta \left\| \bbeta \right\| _{1}, 
\end{equation}
where $\eta>0$ is the regularization parameter, and the empirical loss function is
\begin{equation}\label{loss_func}
 {\mathcal{L}}_{\tau,i}(\bbeta)=\left\| l_{\tau}\left( \mathcal{Y}_{i}- \mathcal{X}_{i} \bbeta \right)/k_n \right\| _{1}.
\end{equation}
In Theorem \ref{Thm1}, we show that the proposed instantaneous beta estimator $ \hat{\bbeta}_{i \Delta_n}$ is consistent with appropriate $\tau$ and $\eta$.
Then, we can estimate the integrated beta using the integration of $\hat{\bbeta}_{i \Delta_n}$'s.
However, their integration cannot enjoy the law of large numbers property since each $\hat{\bbeta}_{i \Delta_n}$ is biased due to the regularization term.
That is, the error of their integration is dominated by the bias terms, which leads to the same convergence rate as that of $\hat{\bbeta}_{i \Delta_n}$. 
Thus, to reduce the effect of the bias and obtain a faster convergence rate, we propose a debiasing scheme as follows.
First, we estimate the inverse instantaneous volatility matrix at time $i \Delta_n$, $\bOmega(i \Delta_n)=\bSigma^{-1}(i \Delta_n)$, where $\bSigma(t)=\bsigma(t)\bsigma^{\top}(t)$.
Specifically, we use the following constrained $\ell_1$-minimization for inverse matrix estimation (CLIME) \citep{cai2011constrained}:
\begin{equation} \label{CLIME}
	\hat{\bOmega}_{i \Delta_n}= \arg \min \| \bOmega\|_1 \quad \text{s.t.} \quad  \|   \frac{1}{k_n \Delta_n }    \mathcal{X}_i  ^{\top}\mathcal{X} _i \bOmega - \bI  \|_{\max} \leq \lambda, 
\end{equation}
where $\lambda$ is the tuning parameter, which will be specified in Theorem \ref{Thm2}. 
With the inverse volatility matrix estimator $\hat{\bOmega}_{i \Delta_n}$, we usually adjust the instantaneous beta estimator $ \hat{\bbeta}_{i \Delta_n}$ as follows:
\begin{equation*}
	\tilde{ \bbeta}_{i \Delta_n}^{\prime} = \hat{\bbeta}_{i \Delta_n}  +  \frac{1}{k_n \Delta_n }   \hat{\bOmega}_{i \Delta_n}^{\top} \mathcal{X}_i  ^{\top}  \left( \mathcal{Y}_i - \mathcal{X}_i  \hat{\bbeta}_{i \Delta_n} \right).
\end{equation*}
The above adjustment is based on the debiasing scheme, which is widely employed in high-dimensional literature \citep{javanmard2014confidence, javanmard2018debiasing, van2014asymptotically, zhang2014confidence}. 
This scheme reduces the bias from $\ell_1$-regularization by adding a bias-correction term to the original estimator, where the correction uses an approximate inverse volatility matrix of the covariates to re-weight the residuals.
Specifically, it examines how much each parameter was shrunk by the $\ell_1$-penalty and then adjusts the estimates to remove that excess shrinkage.
We note that this debiasing scheme performs well under the sub-Gaussian assumption \citep{javanmard2014confidence, javanmard2018debiasing, Kim2024regression, van2014asymptotically}.
However, $\Delta_{i}^{n}Z^c$ has only finite $\(2+\zeta\)$-th moment for an arbitrarily small $\zeta>0$; thus, the debiased instantaneous beta estimator has heavy-tails.
To handle this issue, we employ the Winsorization method as follows.
Define the truncation (Winsorization) function
\begin{equation*}
\psi_{\varpi}\left( x\right) =\begin{cases}x & \text{ if } \left| x\right| \leq \varpi \\  \mbox{sign}(x)\varpi & \text{ if } \left| x\right| > \varpi, \end{cases}
\end{equation*}
where $\varpi > 0$ is a truncation parameter and denote $\psi_{\varpi}\left(\bx \right) =\left(\psi_{\varpi}\left( x_1\right), \ldots, \psi_{\varpi}\left( x_{p_1}\right)\right)^{\top}$ for any vector $\bx=\left(x_1, \ldots, x_{p_1}\right)^{\top}\in \mathbb{R}^{p_{1}}$.
Using this truncation function, we adjust $\hat{\bbeta}_{i \Delta_n}$ as  
\begin{equation}\label{debias}
	\tilde{ \bbeta}_{i \Delta_n} = \hat{\bbeta}_{i \Delta_n}  +  \psi_{\varpi}\left(\frac{1}{k_n \Delta_n }   \hat{\bOmega}_{i \Delta_n}^{\top} \mathcal{X}_{(i+k_n)}  ^{\top}  \left( \mathcal{Y}_{(i+k_n)} - \mathcal{X}_{(i+k_n)}  \hat{\bbeta}_{i \Delta_n} \right)\right),
\end{equation}
where the truncation parameter $\varpi$ will be specified in Theorem \ref{Thm2}.
We note that for the debiasing step, we use the non-overlapping window for  $\mathcal{X}$ and $\mathcal{Y}$, which helps enjoy the martingale property.
Specifically, since $\bbeta_0((i+k_n)\Delta_n)-\hat{\bbeta}_{i \Delta_n}$ is measurable at time $(i+k_n)\Delta_n$, we can handle the noises from $\mathcal{X}_{(i+k_n)}$ and $\mathcal{Y}_{(i+k_n)}$ using the martingale convergence theorem.
We also note that the purpose of the debiasing is to enjoy the law of large numbers property when obtaining the integrated beta estimator.
Usually, the debiasing scheme is employed to obtain asymptotic normality, which enables the hypothesis test or confidence interval construction \citep{javanmard2014confidence, javanmard2018debiasing, van2014asymptotically, zhang2014confidence}.
However, in this paper, we do not focus on this issue and mainly focus on the integrated beta estimation.
Then, the integrated beta estimator is defined as follows:
\begin{equation}\label{Inte}
	\hat{I \beta}  = \sum_{i=0}^{[1/(k_n \Delta_n) ]-2}\tilde{\bbeta}_{i k_n \Delta_n} k_n \Delta_n.
\end{equation}
The debiased LASSO integrated beta estimator $\hat{I \beta}$ can achieve a faster convergence rate than the simple integration of the instantaneous beta estimators.
However, due to the bias adjustment term, it cannot account for the sparsity structure of the integrated beta.
To accommodate the sparsity, we employ the following thresholding scheme:
\begin{equation*}
	\tilde{I\beta}_i= s (\hat{I\beta}_i)  \1 \(  |\hat{I\beta}_i  | \geq h_n  \) \quad \text{and} \quad  \tilde{I \beta} = \( \tilde{I\beta}_i \)_{i=1,\ldots,p},
\end{equation*}
where the thresholding function $s( \cdot)$ satisfies $|s (x)-x| \leq h_n$ and $h_n$ is a thresholding level, which will be specified in Theorem \ref{Thm3}. 
For example, we can employ the hard thresholding function $s(x)=x$ or soft thresholding function $s(x)= x- \mbox{sign}(x) h_n$. 
In the empirical study,  we used the hard thresholding function $s(x)=x$.
We call this the Robust thrEsholding Debiased LASSO (RED-LASSO) estimator. We describe the RED-LASSO estimation procedure in Algorithm \ref{RED-LASSO}.

\begin{algorithm}[H]
         \caption{RED-LASSO estimation procedure}   \label{RED-LASSO}
         \begin{algorithmic}
 \State \textbf{Step 1 } Obtain the instantaneous beta estimator:
 \begin{equation*}  
 \hat{\bbeta}_{i \Delta_n} = \arg \min_{\bbeta \in \mathbb{R}^{p}}\left\| l_{\tau}\left( \mathcal{Y}_{i}- \mathcal{X}_{i} \bbeta \right)/k_n \right\| _{1} + \eta \left\| \bbeta \right\| _{1},
\end{equation*}
 where $\tau=C_{\tau} n^{-1/4} (\log p)^{-3/4}$, $\eta=C_{\eta}\Big[s_p n^{-5/4}\sqrt{\log p} + n^{-5/4} (\log p)^{3/4}\Big]$, and $k_n = c_{k} n^{1/2}$ for some large constants  $C_{\tau}$, $C_\eta$, and $c_{k}$.
 
 \State  \textbf{Step 2 }  Obtain the inverse instantaneous volatility matrix estimator: 
 \begin{equation*}  
	\hat{\bOmega}_{i \Delta_n} = \arg \min \| \bOmega\|_1 \quad \text{s.t.} \quad  \|   \frac{1}{k_n \Delta_n }    \mathcal{X}_{i}  ^{\top}\mathcal{X} _{i} \bOmega - \bI  \|_{\max} \leq \lambda, 
\end{equation*}
where $\lambda=C_\lambda n^{-1/4} \sqrt{\log p} $    for some large constant $C_\lambda$.

 \State  \textbf{Step 3 } Debias the instantaneous beta estimator:
 \begin{equation*}
	\tilde{ \bbeta}_{i \Delta_n} = \hat{\bbeta}_{i \Delta_n}  +  \psi_{\varpi}\left(\frac{1}{k_n \Delta_n }   \hat{\bOmega}_{i \Delta_n}^{\top} \mathcal{X}_{(i+k_n)}  ^{\top}  \left( \mathcal{Y}_{(i+k_n)} - \mathcal{X}_{(i+k_n)}  \hat{\bbeta}_{i \Delta_n} \right)\right),
\end{equation*}
where  $\varpi = C_{\varpi}s_p^{2-\delta}n^{\delta/4}(\log p)^{(1-3\delta)/4}$  for some large constant $C_{\varpi}$.

\State  \textbf{Step 4 }  Obtain the integrated beta estimator:
\begin{equation*}
	\hat{I \beta}  = \sum_{i=0}^{[1/(k_n \Delta_n) ]-2}\tilde{\bbeta}_{i k_n \Delta_n} k_n \Delta_n.
	\end{equation*}

\State   \textbf{Step 5 } Threshold the integrated beta estimator:
	\begin{equation*}
	\tilde{I\beta}_i= s (\hat{I\beta}_i)  \1 \(  |\hat{I\beta}_i  | \geq h_n  \) \quad \text{and} \quad  \tilde{I \beta} = \( \tilde{I\beta}_i \)_{i=1,\ldots,p},
\end{equation*}
where  $s (\cdot)$ satisfies $|s (x)-x| \leq h_n$,  $h_n= C_h b_n$ for some large constant $C_h$, and $b_n$ is defined in Theorem \ref{Thm2}.
\end{algorithmic}
\end{algorithm}

 \subsection{Theoretical results}\label{Theory}
In this section, we investigate asymptotic properties of the proposed RED-LASSO estimation procedure.
To investigate the theoretical properties, we make the following assumptions.
  \begin{assumption}\label{assumption1}~
  \begin{itemize}
 
  \item [(a)] The residual process $Z^c(t)$ and jump size processes, $J^y(t)$ and $\bJ(t)=\(J_1(t), \ldots, J_p(t)\)^{\top}$, satisfy
   \begin{eqnarray*}
   && \max_{1 \leq i \leq n} \mathbb{E} \left\{ |\sqrt{n} \Delta_{i}^{n}Z^c|^{\gamma} \Big| \FF_{(i-1)\Delta_n} \right\} \leq C, \cr
   &&  \sup_{0 \leq t \leq 1}\mathbb{E}\{ |J^y(t)|^{\gamma} \} \leq C, \quad \text{ and } \quad \sup_{0 \leq t \leq 1} \max_{1\leq i \leq p}\mathbb{E}\{ |J_{i}(t)|^{\gamma} \} \leq C  \text{ a.s.},
 	\end{eqnarray*}
 	where $\gamma = 2+\zeta$ for an arbitrarily small $\zeta>0$.
 	
 \item [(b)] The processes  $\bmu(t)$,  $\bmu_{\beta}(t)$, $\bbeta(t)$,  $\bSigma(t)$, and $\bSigma_{\beta}(t) = \bnu_{\beta}(t)\bnu_{\beta}^{\top}(t)$ are almost surely entry-wise bounded, and $ \| \bSigma^{-1}(t)\|_1 \leq C $  \text{ a.s.}
 
 \item [(c)] The processes $\bmu_{\beta}(t) = \(\mu_{\beta,1}(t), \ldots, \mu_{\beta,p}(t) \)^{\top}$ and $\bSigma_{\beta}(t)=\(\Sigma_{\beta, ij}(t)\)_{i,j=1, \ldots, p}$  satisfy the following sparsity condition for $\delta \in [0, 1)$: 
 \begin{equation*}
 \sup_{0 \leq t \leq 1}\sum_{i=1}^p    |\mu_{\beta, i}(t) |^{\delta}   \leq s_p   \quad \text{and} \quad  \sup_{0 \leq t \leq 1}\sum_{i=1}^p |\Sigma_{\beta, ii}(t)|^{\delta/2}    \leq s_p  \text{ a.s.}
 \end{equation*}

 \item[(d)] $ n^{c_1}  \leq p \leq c_2 \exp ( n^{c_3}) $ for some positive constants  $c_1$, $c_2$, and $c_3<1/6$, and $ s_p^2 \log p  \Delta_n k_n  \rightarrow 0$ as $n,p \rightarrow \infty$. 
 
\item[(e)]  Define $\mathcal W_{t} = \left\{ \bw \in \mathbb{R}^{p}:\text{ }\left\| \bw_{S_{t}^{c}} \right\|_{1} \leq 3\left\| \bw_{S_{t}}\right\|_{1} + 4\left\| (\bbeta_{0}(t))_{S_{t}^c}\right\|_{1} \right\}$, where
$\bw_{S_{t}^{c}}$ is the subvector obtained by stacking $\left\{\bw_{j}: \text{ } j \in S_{t}^{c} \right\}$,
$\bw_{S_{t}}$ is the subvector obtained by stacking $\left\{\bw_{j}: \text{ } j \in S_{t} \right\}$, 
$ (\bbeta_{0}(t))_{S_{t}^c}$ is the subvector obtained by stacking $\left\{(\bbeta_{0}(t))_{j}: \text{ } j \in S_{t}^c \right\}$, and $S_{t} = \{j: \text{ jth element}$ \\ of $| \bbeta_0(t) | > n \eta \}$.
Then, there exists a positive constant $\kappa$ such that the following inequality holds for some $D = (8+48/ \kappa)s_{p}(n\eta)^{1-\delta}$ and $0 \leq i \leq n-k_{n}$, where the specific value of $\eta$ is given in Theorem \ref{Thm1}:
\begin{equation*}
\inf\{ \bw^{\top}\nabla ^{2}{\mathcal{L}}_{\tau,i}(\bbeta) \bw: \text{ } \bw \in \mathcal W_{i \Delta_n}, \text{ } \left\| \bw\right\| _{2}= 1,\text{ } \left\| \bbeta -\bbeta_{0}(i \Delta_n)\right\| _{1}\leq D \} \geq \kappa/n .
\end{equation*}

  \item [(f)] The volatility  process  $\bSigma(t) = (\Sigma_{ij}(t) ) _{i,j=1,\ldots, p}$  satisfies the following condition:
  \begin{equation*}
  	  | \Sigma_{ij}(t) -\Sigma_{ij}(s) | \leq  C \sqrt{|t-s| \log p}     \,  \text{ a.s.}
 \end{equation*}

\end{itemize} 
  \end{assumption}

\begin{remark}
Assumption \ref{assumption1}(a) is the finite $\(2+\zeta\)$-th moment condition for an arbitrarily small $\zeta>0$, which allows the dependent process $Y(t)$, covariate process $\bX(t)$, and residual process $Z^c(t)$ to have heavy-tailed distributions. 
We note that in financial applications, the finite second moment condition is not restrictive \citep{cont2001empirical, gabaix2003theory}.
We also note that the moment condition for $Z^c(t)$ is satisfied when $\Delta_{i}^{n}Z^c$ is an independent random variable and $\mathbb{E} \left\{ |\Delta_{i}^{n}Z^c|^{\gamma} \right\} \leq Cn^{-\gamma/2}$, or $\sup_{0 \leq t \leq 1}\sup_{t \leq s \leq 1}\mathbb{E}\left\{ |\nu(s)|^{\gamma}\Big| \FF_{t} \right\} \leq C \text{ a.s.}$
The boundedness condition Assumption \ref{assumption1}(b) implies the sub-Gaussianity for the continuous part of the covariate process, $\bX^c(t)$, and target parameter, $\bbeta(t)$, which are often required to investigate high-dimensional inferences.
However, the boundedness condition can be relaxed to the local boundedness condition by Lemma 4.4.9 in \citet{jacod2012discretization}.
Specifically, if the asymptotic result, such as stable convergence in law or convergence in probability, is satisfied under the boundedness condition, it is also satisfied under the local boundedness condition.
We note that the local boundedness condition is usually satisfied in financial data.
On the other hand, for the continuous-time regression model, we usually assume that the smallest eigenvalue of $\bSigma(t)$ is bounded from below, which implies that the largest eigenvalue of $\bSigma^{-1}(t)$ is bounded.
In this point of view, the condition $ \| \bSigma^{-1}(t)\|_1 \leq C \, \text{ a.s.}$  is not restrictive.
Even if this condition is replaced by the sparsity condition  $\sup_{0 \leq t \leq 1} \max_{ 1 \leq i \leq p} \sum_{j=1}^p |\omega_{ij}(t) | ^{q} \leq s_{\omega, p } \, \text{ a.s.}$, where $\bSigma^{-1}(t) = (\omega_{ij}(t))_{i,j=1,\ldots,p}$, and $q \in [0, 1)$  and  $s_{\omega, p }$ are the sparsity related variables, the difference in theoretical results is up to $s_{\omega, p }$ order.
Assumption \ref{assumption1}(c) is the sparsity condition for the beta process, which is required to investigate the discretization error when estimating instantaneous betas. 
The sparsity assumptions are well established in financial modeling, where only a small number of factors explain the asset returns.
In Assumption \ref{assumption1}(d), we allow the dimension $p$ to grow with the number of observations $n$ exponentially, which accommodates high-dimensional data settings.
Assumption \ref{assumption1}(e) is the eigenvalue condition for the Hessian matrix $\nabla ^{2}{\mathcal{L}}_{\tau,i}(\bbeta)$, which is called the localized restricted eigenvalue ($LRE$) condition \citep{fan2018lamm, sun2020adaptive}. This implies strictly positive restricted eigenvalues over a local neighborhood.
In high-dimensional applications, global Restricted Eigenvalue (RE) conditions are often restrictive due to strong correlations among covariates. 
By focusing on local neighborhoods around the true parameter, the LRE condition provides a more realistic assumption for real data.
We note that $n \eta$ converges to zero for the choice of $\eta$ in Theorems \ref{Thm1}--\ref{Thm2}.
When the coefficient process $\bbeta(t)$ satisfies the exact sparsity condition, i.e., $\delta=0$,  $\mathcal W_{t}$ is replaced  by a $\ell_1$-cone $\left\{ \bw \in \mathbb{R}^{p}:\text{ }\left\| \bw_{S_{t}^{c}} \right\|_{1} \leq 3\left\| \bw_{S_{t}}\right\|_{1}\right\}$, where $S_{t} = \{j: \text{ jth element of } \bbeta_0(t) \neq 0 \}$.
Finally, we need the continuity condition Assumption \ref{assumption1}(f) to investigate asymptotic behaviors of the CLIME estimator. 
We note that this condition is obtained with high probability when $\bSigma(t)$ follows a continuous It\^o diffusion process with bounded drift and instantaneous volatility processes.
We also note that this condition is generally reasonable in financial markets, except the cases of volatility spikes.
However, such spikes can be separately modeled as jumps and handled under the finite activity assumption.
\end{remark}

The following theorem derives the asymptotic properties of the instantaneous beta estimator $ \hat{\bbeta}_{i \Delta_n}$.
Note that the subscript $0$ represents the true parameters. 

\begin{theorem} \label{Thm1}
Under Assumption \ref{assumption1}(a)--(e), let $k_n = c_k n^{c}$ for some constants $c_k$ and $c \in [3/8,3/4]$. 
For any given positive constant $a$, choose $\tau=C_{\tau,a}\sqrt{k_n \Delta_n} (\log p)^{-3/4}$ and $\eta=C_{\eta,a}\Big[s_p n^{-3/2}\sqrt{k_n \log p}$ \\ $+ n^{-1} k_n^{-1/2}(\log p)^{3/4}\Big]$ for some large constants $C_{\tau,a}$ and $C_{\eta,a}$. 
Then, we have, for large $n$,
\begin{equation}\label{Thm1-result1}
       \max_{i} \| \hat{\bbeta}_{i \Delta_n} - \bbeta_{0}(i \Delta_n) \|_{1} \leq  C s_p (n\eta)^{1-\delta} \quad \text{and}  \quad \max_{i} \| \hat{\bbeta}_{i \Delta_n} - \bbeta_{0}(i \Delta_n) \|_{2} \leq  C \sqrt{s_p} (n \eta)^{1- \delta/2},  
\end{equation}
with probability greater than $1-p^{-a}$. 
\end{theorem}

\begin{remark}
Theorem \ref{Thm1} shows the  $\ell_1$ and $\ell_2$ norm error bounds of the instantaneous beta estimator.
We note that as $k_n$ increases, the statistical estimation error decreases, and the time variation approximation error increases.
To achieve the optimality, we choose $c=1/2$, which implies that these two errors have the same convergence rate.
Then, the instantaneous beta estimator has the $\ell_1$ convergence rate of $n^{-(1-\delta)/4}$ and  $\ell_2$ convergence rate of $n^{-(2-\delta)/8}$  with the $\log$ order and sparsity level terms.  
\end{remark}

To estimate the integrated beta, we can use the integration of the instantaneous beta estimators.
However, as discussed in Section \ref{SEC-Estimation}, it cannot enjoy the law of large numbers property due to the heavy-tailed biases.
To tackle this problem, we employ the robust debiasing method \eqref{debias} and obtain the debiased LASSO integrated beta estimator $\hat{I\beta}$ in \eqref{Inte}.
The following theorem establishes the asymptotic behaviors of $\hat{I\beta}$.

\begin{theorem} \label{Thm2}
Under the assumptions in Theorem \ref{Thm1} and Assumption \ref{assumption1}(f), choose $k_n=  c_k n^{1/2}$ for some constant $c_k$.  For any given positive constant $a$, let $\lambda=C_{\lambda,a} n^{-1/4} \sqrt{\log p}$ and $\varpi = C_{\varpi}s_p^{2-\delta}n^{\delta/4}(\log p)^{(1-3\delta)/4}$ for some constants $C_{\lambda,a}$ and $C_{\varpi}$.
Then, we have, with probability greater than $1-p^{-a}$,
\begin{equation}\label{Thm2-result1}
 \|  \hat{I\beta}  - I\beta_0 \| _{\max}  \leq C  b_n,
\end{equation}
where $b_n=   s_p^{2-\delta}n^{(-2+\delta)/4}(\log p)^{(5-3\delta)/4} + s_p n^{-1/2}\(\log p\)^{3/2}$.
\end{theorem}

\begin{remark}
Theorem \ref{Thm2} shows the max norm error bound of the debiased LASSO integrated beta estimator.
When the beta process satisfies the exact sparsity condition, i.e., $\delta=0$,  the debiased LASSO  integrated beta estimator has the convergence rate of $s_p^2 n^{-1/2}\(\log p\)^{5/4}+s_p n^{-1/2}\(\log p\)^{3/2}$, while we have a slower convergence rate of $s_p^2 n^{-1/4}\sqrt{\log p}+s_p n^{-1/4}\(\log p\)^{3/4}$ without a debiasing scheme.
The $n^{1/2}$ term is the optimal convergence rate of estimating model parameters given $n$ observations.
For the $\log$ order term, the usual optimal rate is $\sqrt{\log p}$ in high-dimensional inferences.
However, we have $\(\log p\)^{3/2}$ term since the additional $\log p$ term comes from bounding the time-varying processes, such as the target process $\bbeta(t)$.
In sum, the debiased LASSO integrated beta estimator has the optimal convergence rate with up to $\log p$ and $s_p$ orders. 
\end{remark}

Theorem \ref{Thm2} reveals that  the debiased LASSO integrated beta estimator performs better than the integration of the instantaneous beta estimators.
Finally, to account for the sparsity structure, we threshold the debiased LASSO integrated beta estimator and obtain the RED-LASSO estimator.
Theorem \ref{Thm3} establishes the $\ell_1$ convergence rate of the RED-LASSO estimator.

 \begin{theorem} \label{Thm3}
Under the assumptions in Theorem \ref{Thm2}, for any given positive constant $a$, choose $h_n= C_{h,a} b_n$ for some constant $C_{h,a}$.
Then, we have, with probability greater than $1-p^{-a}$,
\begin{equation}\label{Thm3-result1}
 \|  \tilde{I\beta}  - I\beta_0 \| _{1}  \leq C    s_p  b_n^{1-\delta}.
\end{equation}
\end{theorem}

Theorem \ref{Thm3} shows that the proposed RED-LASSO estimator is consistent in terms of the $\ell_1$ norm.
We note that under the sub-Gaussian assumption on the log-return data, \citet{Kim2024regression} proposed the integrated beta estimator that has the $\ell_1$ convergence rate of $s_p a_n^{1-\delta}$, where $a_n=s_p^{2-\delta}n^{(-2+\delta)/4}(\log p)^{(2-\delta)/2} +s_p s_{\omega,p}n^{(-2+q)/4}(\log p)^{(2-q)/2}  + s_p n^{-1/2}\(\log p\)^{3/2}$, and $s_{\omega,p}$ and $q$ are the sparsity related terms for the inverse volatility matrix.
Thus, the cost of handling the heavy-tailedness is at most $\log p$ order.
We also note that the regularized approximate quadratic (RA-Lasso) estimator \citep{fan2017estimation} and the regularized adaptive Huber estimator \citep{sun2020adaptive} are robust to heavy-tailed distributions.
Under the constancy of the beta process, these estimators achieve the $\ell_1$ convergence rate of $s_p n^{-1/2} \sqrt{\log p}$.
In contrast, the proposed RED-LASSO estimator accommodates both heavy-tailedness and time-varying beta processes, with an additional cost of at most $\log p$ and $s_p$ orders.

\subsection{Discussion on the tuning parameter selection}\label{SEC-Tuning}
In this section,  we discuss how to choose the tuning parameters to implement the RED-LASSO estimation procedure.
We first obtain the variables $\Delta_{i} ^n \hat{X}^c_j$, $j=1, \ldots, p$, based on the threshold level \eqref{jump_adj}.
Then, to handle the scale problem, we standardize the variables $\Delta_{i} ^n {Y}$ and $\Delta_{i} ^n \hat{X}^c_j$, $j=1, \ldots, p$,  to have a zero mean and unit variance.
The re-scaling is employed after obtaining the RED-LASSO estimator.
In the local regression stage \eqref{ins_beta}, we select $k_{n}=[n^{1/2}]$ for the simulation study.
The choice of $k_n$ for the empirical study is presented in Section \ref{SEC-5}.
Also, we choose 
\begin{eqnarray}\label{tuning} 
&&\tau = c_{\tau}n^{-1/4}\(\log p\)^{-3/4} , \quad \eta = c_{\eta}n^{-5/4}\(\log p\)^{3/4}, \cr
&& \lambda = c_{\lambda}n^{-1/4}\sqrt{\log p}, \quad \varpi = c_{\varpi}\(\log p\)^{1/4}, \quad \text{ and } \quad h_n=c_h n^{-1/2} \(\log p\)^{3/2},
\end{eqnarray}
where $c_{\tau}$, $c_{\eta}$, $c_{\lambda}$, $c_{\varpi}$, and $c_h$ are tuning parameters.
For the simulation and empirical studies, we choose $c_{\varpi}$ and $c_h$ that minimize the corresponding mean squared prediction error (MSPE).
The results are $c_{\varpi}=0.025$, and $c_h=0.1$.
Details can be found in Section \ref{SEC-5}.
Also,  we select $c_{\tau}, c_{\eta} \in [0.1, 10]$ via 5-fold cross-validation based on the following mean squared error (MSE):
\begin{equation*}
 \frac{1}{k_{n}^{\text{test}}} \left\|\mathcal{Y}_{i}^{\text{test}} - \(\mathcal{X}_{i}^{\text{test}}\) \hat{\bbeta}_{i \Delta_n}^{\text{train}}\right\|_2^2
\end{equation*}
where $\mathcal{Y}_{i}^{\text{test}}$ and  $\mathcal{X}_{i}^{\text{test}}$ are obtained from $\mathcal{Y}_{i}$ and $\mathcal{X}_{i}$, respectively, by selecting the rows corresponding to the test data, $k_{n}^{\text{test}}$ is the number of rows in $\mathcal{Y}_{i}^{\text{test}}$ and $\mathcal{X}_{i}^{\text{test}}$, and $\hat{\bbeta}_{i \Delta_n}^{\text{train}}$ is the instantaneous beta estimator calculated from the training data.
The test and training data are chosen based on the 5-fold cross-validation.
This choice procedure is similar to that of \citet{sun2020adaptive, tan2023sparse}, which employ robust regularization approaches.
Similarly, we choose  $c_{\lambda} \in [0.1, 10]$ via 5-fold cross-validation based on the following loss function \citep{cai2011constrained}:
\begin{equation*}
\tr\[\( \frac{1}{k_{n}^{\text{test}} \Delta_n }    \(\mathcal{X}_{i}^{\text{test}}\)^{\top} \mathcal{X}_{i}^{\text{test}} \hat{\bOmega}_{i \Delta_n}^{\text{train}} - \bI_{p}\)^2\],
\end{equation*}
where $\hat{\bOmega}_{i \Delta_n}^{\text{train}}$ is the inverse instantaneous volatility matrix estimator obtained from the training data and $\bI_{p}$ is the $p$-dimensional identity matrix.
We note that  $\hat{\bbeta}_{i \Delta_n}^{\text{train}}$ and $\hat{\bOmega}_{i \Delta_n}^{\text{train}}$ enjoy the same theoretical properties as  $\hat{\bbeta}_{i \Delta_n}$ and $\hat{\bOmega}_{i \Delta_n}$, which can be shown similar to the proofs of Theorems \ref{Thm1}--\ref{Thm2}.
This holds because the time gaps used in each local estimation are sufficiently short, which allows us to control the errors introduced by subsampling.

\section{A simulation study} \label{SEC-4}

To check the finite sample performance of the proposed RED-LASSO estimator, we conducted simulations.
Based on the models \eqref{model-1}--\eqref{model-2}, we generated the data using the heavy-tailed and sub-Gaussian processes with frequency $1/n^{all}$.
Specifically, we employed the following time-series regression jump-diffusion model:
\begin{eqnarray*}
&& dY(t)= \bbeta ^{\top}(t)  d\bX^c(t)+d Z^c(t)+J^y(t) d \Lambda^y(t), \quad d\bX(t)=  d \bX^c(t) + d \bX^J(t), \cr
&&  d\bX^c(t)= \bsigma(t) d\bB(t), \quad  d \bX^J(t) =  \bJ(t) d \bLambda(t), \quad \text{and} \quad    dZ^c(t)= \nu(t) d W(t),
\end{eqnarray*}
where the jump sizes $J_{i}(t)$ and $J^y(t)$ were obtained from 0.1 times i.i.d. $t$-distribution with degrees of freedom $df$, and $\bLambda(t)=\(\Lambda_{1}(t), \ldots, \Lambda_{p}(t) \)^{\top}$ and $\Lambda^y(t)$ were generated by Poisson processes with the intensities $\(20, \ldots, 20\)^{\top}$ and $10$, respectively. We chose $df$ as $2$ and $\infty$ for the heavy-tailed and sub-Gaussian processes, respectively. 
The initial values of $X(t)$ and $Y(t)$ were set as zero, and we generated $\nu(t)$  as follows:
\begin{equation*}
\nu(t_l)=\left(1+0.5\left|t_{df,l}\right|\right)\nu ^{\prime} (t_l),
\end{equation*}
where $t_{df,l}$, $l=1, \ldots, n^{all}$, are the i.i.d. $t$-distributions  with degrees of freedom $df$, and $\nu^{\prime}(t_l)$, $l=1, \ldots, n^{all}$, were generated from the following Ornstein-Uhlenbeck process: 
\begin{equation*}
d\nu^{\prime}(t)=3\(0.8-\nu^{\prime}(t)\)dt + 0.24 d\bW^{\nu}(t),
\end{equation*}
where $\nu^{\prime}(0)=1$ and $\bW^{\nu}(t)$ is an independent Brownian motion.
We note that the process $\nu(t)$ is not realistic.
However, to investigate the effect of the heavy-tailedness of the return process, the structure of  $\nu(t)$ is imposed.
To generate the volatility process $\bsigma(t)$, we first generated the Ornstein-Uhlenbeck process $u(t)$ as follows:
\begin{equation*}
du(t)=5\left(0.45 - u(t) \right)dt + 0.2d\bW^{u}(t),
\end{equation*}
where $u(0)=1$ and $\bW^{u}(t)$ is an independent Brownian motion. Then, we took $\bsigma(t)$ as a Cholesky decomposition of $\bSigma(t)=\(\Sigma_{ij}(t)\)_{1 \leq i,j \leq p}$, where $\Sigma_{ij}(t) = u(t)0.6^{|i-j|}$.
To generate the coefficient $\bbeta(t)$, we considered the exact sparse process, i.e., $\beta_{i}(t)=0$ for $[s_p]+1 \leq i \leq p$.
Specifically, we generated $\bbeta(t)$ as follows:
\begin{equation*}
	d \bbeta(t)= \bmu_{\beta}(t) dt + \bnu_{\beta}(t) d \bW_{\beta}(t),
\end{equation*}
where $\bmu_{\beta}(t)=\left(\mu_{\beta,1}(t), \ldots, \mu_{\beta,p}(t)\right)^{\top}$, $\bnu_{\beta}(t)=\left(\nu_{\beta,i,j}(t)\right)_{1\leq i,j \leq p}$, and $\bW_{\beta}(t)$ is a $p$-dimensional independent Brownian motion. 
For $1 \leq i \leq [s_p]$, the initial value $\beta_i(0)=1$ and $\mu_{\beta,i}(t)=0.1$ for $0 \leq t \leq 1$.
The process $\left(\nu_{\beta,i,j}(t)\right)_{1\leq i,j \leq [s_p]}$ was taken to be $\xi(t)\bI_{[s_p]}$, where $\bI_{[s_p]}$ is the $[s_p]$-dimensional identity matrix and $\xi(t)$ follows the Ornstein-Uhlenbeck process: 
\begin{equation*}
d\xi(t)=3\(0.3-\xi(t)\)dt + 0.1 d\bW^{\xi}(t),
\end{equation*}
where $\xi(0)=0.15$ and $\bW^{\xi}(t)$ is an independent Brownian motion.
We chose $p=100$, $s_p=\log p$, $n^{all}=4000$, and we varied $n$ from $1000$ to $4000$. 
When implementing the RED-LASSO estimation procedure, the tuning parameters were selected as discussed in Section \ref{SEC-Tuning}.

To investigate the effect of the robustification of the RED-LASSO estimator, we employed a thrEsholding Debiased LASSO (ED-LASSO) estimator.
The ED-LASSO estimator uses the same estimation procedure as the RED-LASSO estimator with $\tau=\varpi=\infty$. 
Since the ED-LASSO estimator does not employ the Huber loss and Winsorization method, the jump adjustment for the dependent process $Y(t)$ is needed. 
Thus, we used $\mathcal{Y}_{i}^{\prime}$ instead of $\mathcal{Y}_{i}$ for the ED-LASSO estimator, where
\begin{eqnarray}\label{jump_Y}
&& \mathcal{Y}_i^{\prime} = 
\begin{pmatrix}
\Delta_{i+1}^n \hat{Y}^c \\ 
\Delta_{i+2}^n \hat{Y}^c \\  
 \vdots \\ 
\Delta_{i+k_n}^n \hat{Y}^c  
\end{pmatrix}
\quad \text{ and } \quad 
\Delta_{i}^n \hat{Y}^c = \Delta_{i} ^n Y \, \1 _{\{|\Delta_{i} ^n Y | \leq u_n \}}.
\end{eqnarray}
In the simulation and empirical studies, we choose  $u_n = \sqrt{BV^Y \log p} n^{-1/2}$, where the bipower variation $BV^Y = \dfrac{\pi}{2}\sum_{i=2}^{n} | \Delta_{i-1} ^n Y| \cdot | \Delta_{i} ^n Y|$.
We note that the ED-LASSO estimator can enjoy the same theoretical properties as the RED-LASSO estimator under the sub-Gaussian process, but it cannot explain the heavy-tailed process.
As a benchmark, we also considered the LASSO estimator \citep{tibshirani1996regression}, which cannot account for any of the heavy-tailed distribution or the time-varying beta process.
Specifically, we employed the LASSO estimator as follows:
\begin{equation}\label{LASSO}
	\tilde{I \beta}^{\lasso}= \argmin _{\bbeta} \left \{ \sum^{n}_{i=1} \(\Delta_{i}^n \hat{Y}^c - \Delta_{i} ^n \hat{\bX} ^{c\top}  \bbeta \)^2 + \eta^{\lasso} \| \bbeta \|_1 \right \},
\end{equation}  
where the regularization parameter $\eta^{\lasso} \in [0.1, 10]$ was selected via 5-fold cross-validation based on the MSE.
We also employed the Support Vector Regression (SVR) estimator with a linear kernel as follows:
\begin{equation}\label{SVR}
	\tilde{I \beta}^{\svr}= \argmin _{\bbeta} \left \{ C_s \sum^{n}_{i=1} \max\left\{ |\Delta_{i}^n \hat{Y}^c - \Delta_{i} ^n \hat{\bX}^{c\top}  \bbeta | -\epsilon , 0\right\} + \dfrac{1}{2} \| \bbeta \|_2^2 \right \},
\end{equation}  
where the cost parameter $C_s \in [10^{-4}, 1]$ and insensitivity parameter $\epsilon \in [10^{-4}, 1]$ were selected by 5-fold cross-validation using the MSE.
We note that the SVR estimator can partially mitigate the effect of heavy-tailed distributions by employing the above epsilon-insensitive loss function instead of a squared loss function.
However, it cannot fully account for heavy-tails, nor can it handle the time-varying property of the beta process.
After obtaining the RED-LASSO, ED-LASSO, LASSO, and SVR estimators, the average estimation errors under the max norm, $\ell_1$ norm, and $\ell_2$ norm were computed by 1000 simulations.

\begin{figure}[!ht]
\centering
\includegraphics[width = 1\textwidth]{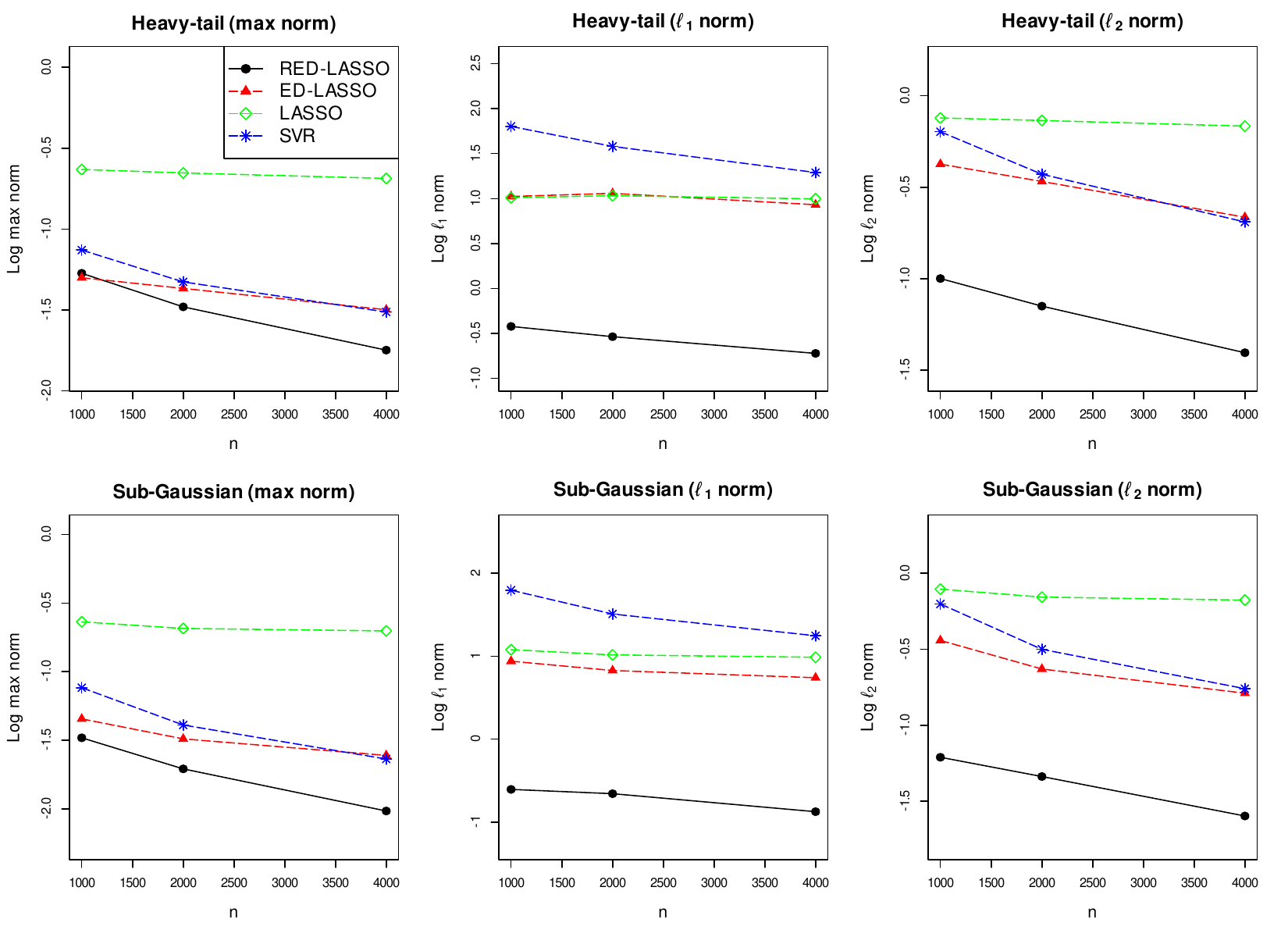}
\caption{The log max, $\ell_1$, and $\ell_2$ norm error plots (corresponding to
three columns) of the RED-LASSO (black dot), ED-LASSO (red triangle), LASSO (green diamond), and SVR (blue star) estimators for $p=100$ and $n=1000, 2000, 4000$.}\label{Beta_simulation}
\end{figure}

Figure \ref{Beta_simulation} plots the log max, $\ell_1$, and $\ell_2$ norm errors of the RED-LASSO, ED-LASSO, LASSO, and SVR estimators with $n=1000, 2000, 4000$ for the heavy-tailed and sub-Gaussian processes.
From Figure \ref{Beta_simulation}, we can find that the estimation errors of the RED-LASSO estimator decrease as the sample size $n$ increases.
As expected, the RED-LASSO estimator performed the best for the heavy-tailed process.
This may be because the RED-LASSO estimator can fully explain the heavy-tailedness, while other estimators cannot.
For the sub-Gaussian process, the RED-LASSO and ED-LASSO estimators showed better performance than the LASSO estimator.
This is because the LASSO estimator cannot account for the time variation of the beta process.
We note that, even for the sub-Gaussian process, the RED-LASSO estimator showed better performance than the ED-LASSO estimator.
One possible explanation for this is that the true return process can have some extreme values over time, even if the sub-Gaussian random variables are used.
From this result, we can conjecture that the RED-LASSO estimator is robust to the heavy-tailedness of the log-return process.
We also note that the ED-LASSO estimator does not outperform the SVR estimator for both heavy-tailed and sub-Gaussian processes.
This may be because, although the SVR estimator cannot account for the time-varying property of the beta process, it handles heavy-tailed distributions better than the ED-LASSO estimator.

\begin{figure}[!ht]
\centering
\includegraphics[width = 1\textwidth]{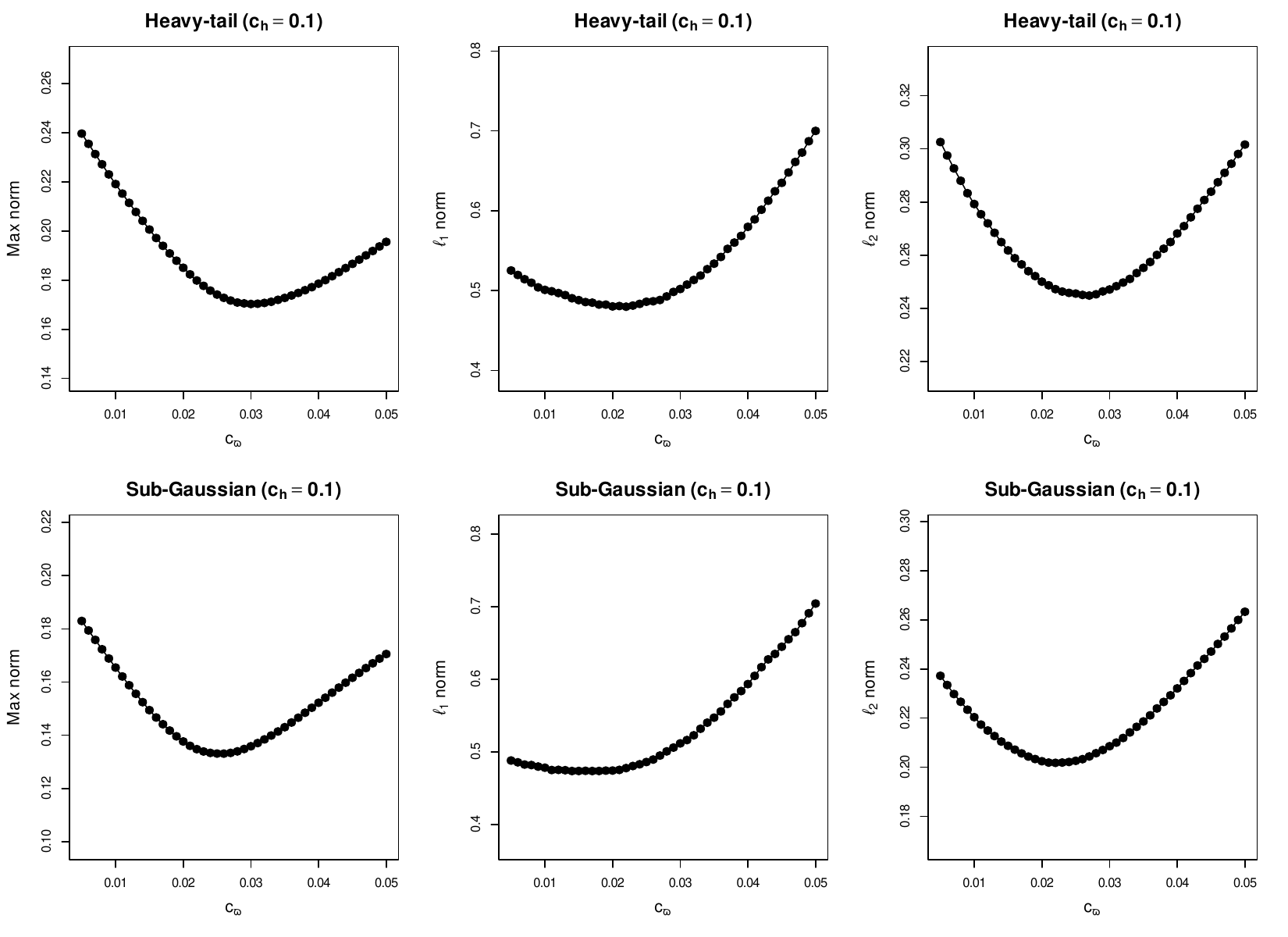}
\caption{The max, $\ell_1$, and $\ell_2$ norm error plots of the RED-LASSO estimator for $p=100$ and $n=4000$ with the heavy-tailed and sub-Gaussian processes. 
Note that $c_{h}=0.1$ is fixed and $c_{\varpi}$ is varied from $0.005$ to $0.05$.}\label{Beta_simulation2}
\end{figure}

\begin{figure}[!ht]
\centering
\includegraphics[width = 1\textwidth]{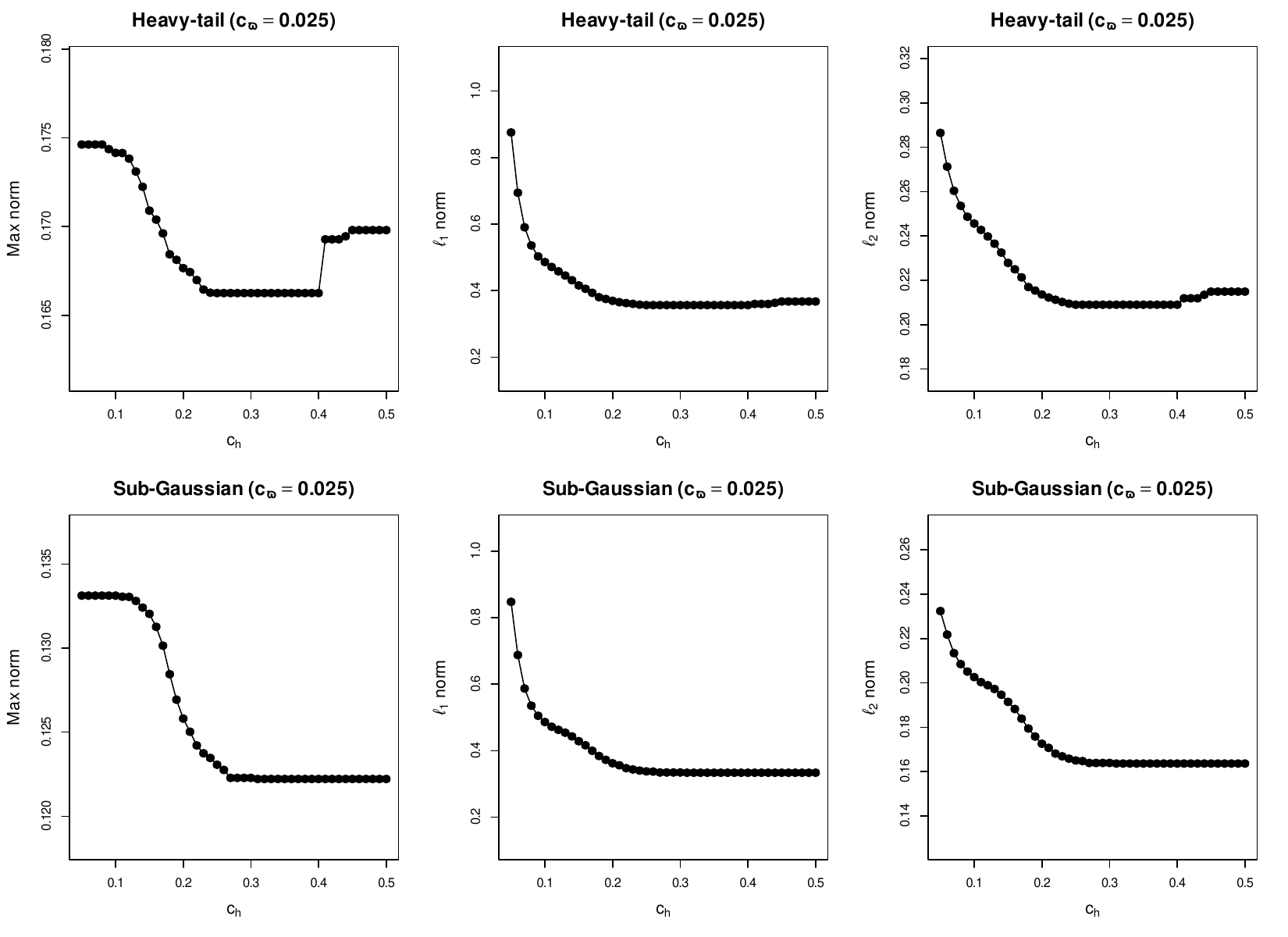}
\caption{The max, $\ell_1$, and $\ell_2$ norm error plots of the RED-LASSO estimator for $p=100$ and $n=4000$ with the heavy-tailed and sub-Gaussian processes. 
Note that $c_{\varpi}=0.025$ is fixed and $c_{h}$ is varied from $0.05$ to $0.5$.}\label{Beta_simulation3}
\end{figure}

On the other hand, we set the tuning parameters $c_{\varpi}=0.025$ and $c_h=0.1$ in the numerical study.
To investigate the robustness of the RED-LASSO estimator with respect to the choice of $c_{\varpi}$ and $c_h$, we calculated the estimation errors by varying $c_{\varpi}$ and $c_h$.
Figures \ref{Beta_simulation2} and \ref{Beta_simulation3} show the max, $\ell_1$, and $\ell_2$ norm errors of the RED-LASSO estimator with $n=4000$ for the heavy-tailed and sub-Gaussian processes.
In Figure \ref{Beta_simulation2}, $c_{h}=0.1$ is fixed and $c_{\varpi}$ is varied from $0.005$ to $0.05$, while $c_{\varpi}=0.025$ is fixed and $c_{h}$ is varied from $0.05$ to $0.5$ in Figure \ref{Beta_simulation3}.
As seen in Figures \ref{Beta_simulation2}--\ref{Beta_simulation3}, the errors of the RED-LASSO estimator usually do not critically depend on the choice of $c_{\varpi}$ and $c_h$.
This result supports the practical applicability of the RED-LASSO procedure.
An exception is the $\ell_1$ norm error for small $c_h$.
This may be because the RED-LASSO estimator with small $c_h$ cannot effectively handle the non-sparse structure of the debiased integrated beta estimator.


\section{An empirical study} \label{SEC-5}
In this section, we applied the proposed RED-LASSO estimator to high-frequency trading data from January 2013 to December 2019.
We took stock price data, futures price data, and firm fundamentals from the End of Day website, FirstRate Data website, and Center for Research in Security Prices (CRSP)/Compustat Merged Database, respectively.
We used 5-minute data to mitigate the effect of microstructure noise.
Specifically, we obtained 5-minute log-price data with the previous tick scheme \citep{wang2010vast, zhang2011estimating} and processed the data similarly to the procedure in \citet{Kim2024regression}.
The days with half trading hours were not included.
For the dependent process, we collected the log-price data of the following five assets: Apple Inc. (AAPL), Berkshire Hathaway Inc. (BRK.B), Amazon.com, Inc. (AMZN), Alphabet Inc. (GOOG), and Exxon Mobil Corporation (XOM). 
These firms have the top market values in their global industry classification standard (GICS) sectors.
For the covariate process, we first obtained the log-prices of 54 futures, which are often used as the market macro variables.
For example, we selected 20 commodity futures data, 10 currency futures data, 10 interest rate futures data, and 14 stock market index futures data.  
The specific list is presented in Table \ref{Table3} in the Appendix.
Then, we constructed Fama-French five factors \citep{fama2015five} and the momentum factor \citep{carhart1997persistence} with the assets listed on NYSE, NASDAQ, and AMEX, which are widely used in stock market analysis. 
We note that MKT, HML, SMB, RMW, CMA, and MOM represent the market, value, size, profitability, investment, and momentum factors, respectively.
First, we calculated MKT as the return of a value-weighted portfolio of whole assets.
Then, we obtained other factors as follows:
\begin{eqnarray*}
&& HML=\(SH+BH\)/2-\(SL+BL\)/2, \cr
&& SMB=\(SH+SM+SL\)/3-\(BH+BM+BL\)/3, \cr
&& RMW=\(SR+BR\)/2-\(SW+BW\)/2, \cr
&& CMA=\(SC+BC\)/2-\(SA+BA\)/2,  \cr
&& MOM=\(SU+BU\)/2-\(SD+BD\)/2, 
\end{eqnarray*}
where small (S) and big (B) portfolios represent the small and big market equities, respectively, while we classified high (H), medium (M), and low (L) portfolios according to their ratio of book equity to market equity.
On the other hand, robust (R), neutral (N), and weak (W) portfolios were classified by their profitability, while we obtained conservative (C), neutral (N), and aggressive (A) portfolios using their investment data. 
Also, up (U), flat (F), and down (D) portfolios were classified by their momentum of the return.
The portfolio constituents were updated monthly, and, with 5-minute frequency, we obtained the portfolio return as follows:
\begin{equation*}
WRet_{d,i}=\dfrac{\sum_{j=1}^{N_d}w_{d,i}^{j} \times Ret_{d,i}^{j}}{\sum_{j=1}^{N_d}w_{d,i}^{j}},
\end{equation*}
where $WRet_{d,i}$ is the portfolio return for the $d$th day and $i$th time interval, $N_d$ is the number of portfolio components on the $d$th day, the superscript $j$ is used to represent the $j$th stock of the portfolio, and $w_{d,i}^{j}$ is calculated by
\begin{equation*}
w_{d,i}^{j}=w_{d}^{j} \times \prod_{l=0}^{i-1}\(1+Ret_{d,l}^{j}\),
\end{equation*}
where $w_{d}^{j}$ is the market capitalization of the $j$th stock at the market close time on the day $d-1$, and $Ret_{d,0}^{j}$
represents the overnight return from the day $\(d-1\)$ to  day $d$. 
To sum up, the five assets and 60 factors were used for the dependent and covariate processes, respectively.
The details of the data processing can be found in \citet{ait2020high} and \citet{Kim2024regression}.

When obtaining the RED-LASSO estimator, we selected the tuning parameters based on Section \ref{SEC-Tuning} and Section \ref{SEC-4}. 
Also, we set $k_n = 78$ so that the instantaneous betas were estimated on a daily basis, while the integrated betas were estimated on a monthly basis.
Finally, we chose the tuning parameters $c_{\varpi}$ and $c_h$ using the cross-validation scheme on multi-period high-frequency financial data in 2013.
Since the integrated beta estimator is obtained at the monthly level and the stationarity assumption is reasonable for the beta process, we employed one-month-ahead prediction error to evaluate performance for each tuning parameter selection \citep{wang2010vast}.
Specifically, we first defined the following mean squared prediction error (MSPE):
\begin{equation*}
\Lambda(c_{\varpi},c_h) = \frac{1}{55} \sum_{s=1}^{5} \sum_{j=1}^{11} \left\| \tilde{I \beta}^{j,s}\(c_{\varpi}, c_h\) - \tilde{I \beta}^{(j+1),s}\(\infty, \infty\) \right\|_2^2,
\end{equation*}
where $\tilde{I \beta}^{j,s}\(c_{\varpi},c_h\)$ is the RED-LASSO estimator with the tuning parameters $c_{\varpi}$ and $c_h$ for the $j$th month in $2013$ and $s$th stock. 
We note that $\tilde{I \beta}^{j,s}\(\infty, \infty\)$ is the RED-LASSO estimator obtained without truncation or thresholding.
Then, we selected $c_{\varpi}$ and $c_h$ by minimizing $\Lambda(c_{\varpi},c_h)$ over $c_{\varpi} \in \left\{ 0.005l \mid 1 \leq l \leq 10, \, l \in \mathbb{Z} \right\}$ and $c_h \in \left\{ 0.05l \mid 1 \leq l \leq 10, \, l \in \mathbb{Z} \right\}$.
The results are $c_{\varpi}=0.025$ and $c_h=0.1$.
Then, using the RED-LASSO, ED-LASSO, LASSO, and SVR estimation procedures, we obtained the monthly integrated betas for each of the five assets.
For the non-trading period, we set the beta estimates as zero.

\begin{table}[!ht]
\caption{The average in-sample and out-of-sample $R^2$ of the RED-LASSO, ED-LASSO, LASSO, and SVR estimators over the five assets.}\label{Table1}
\centering
\scalebox{0.85}{
\begin{tabular}{c c c c c c c c c c c  c c c c c c c c c c}
\hline
\multicolumn{2} {c}{} & \multicolumn{7} {c}{In-sample $R^2$}   \\ \cline{3-9}
\multicolumn{2} {c}{} & \multicolumn{7} {c}{Estimator}   \\ \cline{3-9}
&&		RED-LASSO 	 &&	ED-LASSO       && LASSO   &&  SVR  \\     \hline
whole period      &&   0.233 	      &&   0.228 	     &&    0.197  &&    0.226    	    \\
2013	             &&   0.229 	      &&   0.222 	     &&    0.195  &&    0.225      	    \\
2014 		    &&   0.202          &&   0.195 	     &&    0.165	&&    0.192           \\
2015 		    &&   0.246          &&   0.240 	     &&    0.211	&&    0.238         	\\
2016	             &&   0.223 	      &&   0.221 	     &&    0.190  &&    0.223    	    \\
2017 		    &&   0.180          &&   0.180 	     &&    0.151	&&    0.177            \\
2018 	 	    &&   0.320          &&   0.313 	     &&    0.280	&&    0.307         	\\
2019 		    &&   0.228          &&   0.225	     &&    0.191	&&    0.223            \\  \hline
\multicolumn{2} {c}{} & \multicolumn{7} {c}{Out-of-sample $R^2$}   \\ \cline{3-9}  
\multicolumn{2} {c}{} & \multicolumn{7} {c}{Estimator}   \\ \cline{3-9}
&&		RED-LASSO 	 &&	ED-LASSO       && LASSO   &&  SVR  \\   \hline
whole period      &&   0.215 	      &&   0.206 	     &&    0.190  &&    0.199    	     \\
2014 		    &&   0.183          &&   0.170 	     &&    0.157	&&    0.151            \\
2015 		    &&   0.229          &&   0.217 	     &&    0.205	&&    0.214         	\\
2016	             &&   0.206 	      &&   0.198 	     &&    0.183  &&    0.182    	     \\
2017 		    &&   0.165          &&   0.158 	     &&    0.143	&&    0.160            \\
2018 	 	    &&   0.299          &&   0.290 	     &&    0.267	&&    0.287         	\\
2019 		    &&   0.209          &&   0.204	     &&    0.186	&&    0.202            \\  \hline
\end{tabular}
}
\end{table}

We first compare the performances of the RED-LASSO, ED-LASSO, LASSO, and SVR estimators.
To do this, we calculated the monthly in-sample and out-of-sample $R^2$ with the monthly integrated beta estimates.
We note that since the integrated beta reflects the average effect of covariate movements on the dependent process, the higher in-sample $R^2$ implies a better approximation of this average effect.
This indicates how well the estimator captures the overall time-varying relationship between the dependent and covariate processes.
That is, $R^2$ measures the goodness-of-fit for the proposed time-varying linear model.
However, the in-sample $R^2$ can cause the overfitting issue. 
To overcome this, we use the out-of-sample $R^2$. 
Theoretically, under a stationarity condition of the beta process, the best predictor of the beta process is the one from the previous period. 
From this point of view, the out-of-sample $R^2$ indicates the performance of explaining this average relationship over future data. 
Specifically, from the high out-of-sample $R^2$, we can conjecture that the proposed method can account for the dynamics of the beta process. 
The out-of-sample $R^2$ was calculated using the integrated betas from the previous month, and it was obtained excluding the year 2013 since the tuning parameters were chosen based on the data in 2013.
For each year, we calculated the average $R^2$ across the five assets and twelve months.
Table \ref{Table1} shows the average in-sample and out-of-sample $R^2$ of the RED-LASSO, ED-LASSO, LASSO, and SVR estimators.
As seen in Table  \ref{Table1}, the RED-LASSO estimator shows the best performance for all periods.
This may be because only the RED-LASSO estimator can handle both the heavy-tailed distribution of the return process and time-varying property of the beta process.

\begin{table}[!ht]
\scalebox{0.52}{
\begin{tabular}{ll cccccccccccccccccccccccc}
\hline
\multicolumn{1}{l}{Type} && \multicolumn{4}{c}{AAPL} && \multicolumn{4}{c}{BRK.B} && \multicolumn{4}{c}{AMZN} && \multicolumn{4}{c}{GOOG} && \multicolumn{4}{c}{XOM} \\   \cline{3-6} \cline{8-11} \cline{13-16} \cline{18-21} \cline{23-26}  &&  \text{RED} & \text{ED} & \text{LASSO} & \text{SVR} && \text{RED} & \text{ED} & \text{LASSO} & \text{SVR} && \text{RED} & \text{ED} & \text{LASSO} & \text{SVR} && \text{RED} & \text{ED} & \text{LASSO} & \text{SVR} && \text{RED} & \text{ED} & \text{LASSO} & \text{SVR} \\ \hline
All factors 
&& 0.18 & 0.42 & 0.30 & 0.95 && 0.24 & 0.44 & 0.50 & 0.95 && 0.21 & 0.44 & 0.39 & 0.95 && 0.21 & 0.43 & 0.35 & 0.95 && 0.25 & 0.47 & 0.59 & 0.95 \\
Commodity
&& 0.05 & 0.33 & 0.12  & 0.96 && 0.05  & 0.32 & 0.31 & 0.96 && 0.05 & 0.34 & 0.21 & 0.96 && 0.05 & 0.33 & 0.19 & 0.96 && 0.14 & 0.39 & 0.46 & 0.96  \\
Currency
&& 0.08 & 0.39 & 0.18 & 0.97 && 0.10 & 0.40 & 0.40 & 0.97 && 0.11 & 0.41 & 0.30 & 0.97 && 0.12 & 0.39 & 0.23 & 0.97 && 0.12 & 0.41 & 0.49 & 0.97 \\
Interest rate 
&& 0.06 & 0.29 & 0.15  & 0.84 && 0.08  & 0.29 & 0.38 & 0.84 && 0.07  & 0.29 & 0.25 & 0.84 && 0.07 & 0.31 & 0.20 & 0.84 && 0.06 & 0.30 & 0.38 & 0.84 \\
Stock market index 
&& 0.50 & 0.62 & 0.70 & 0.99 && 0.58 & 0.63 & 0.85 & 0.99 && 0.52 & 0.62 & 0.74 & 0.99 && 0.53 & 0.63 & 0.74 & 0.99 && 0.42 & 0.60 & 0.86 & 0.99 \\
Market factor 
&& 0.25 & 0.55 & 0.37 & 1.00 && 0.61 & 0.78 & 0.71 & 1.00 && 0.39 & 0.61 & 0.52 & 1.00 && 0.37 & 0.60 & 0.46 & 1.00 && 0.79 & 0.87 & 0.86 & 1.00 \\
\hline
\end{tabular}
\caption{The average proportion of non-zero monthly integrated beta estimates across factor groups and assets for the RED-LASSO (RED), ED-LASSO (ED), LASSO, and SVR estimators.}
\label{Table2}
}
\end{table}


Table \ref{Table2} reports the monthly average proportion of non-zero integrated beta estimates across six factor groups and five assets over 84 months for the RED-LASSO, ED-LASSO, LASSO, and SVR estimators.
Detailed non-zero frequencies for each individual factor are presented in Table \ref{Table4} in the Appendix.
As seen in Table \ref{Table2}, the RED-LASSO estimator can better account for the sparsity of the integrated betas than the ED-LASSO, LASSO, and SVR estimators.
From this result, we can conjecture that the proposed RED-LASSO provides more sparse beta estimates, which is an important property in practice.
Furthermore, as discussed above, the RED-LASSO estimator shows the best performance in terms of  $R^2$ in Table \ref{Table1}.
That is, the RED-LASSO estimator can explain the market dynamics well with a simpler model.
We note that for the RED-LASSO estimates, the stock market index futures factors had non-zero integrated betas more often than the other futures factors.
This result is consistent with the multi-factor models \citep{asness2013value, carhart1997persistence, fama1992cross, fama2015five} since the market factors can be partially explained by the stock market index futures factors.

\begin{figure}[!ht]
\centering
\includegraphics[height=1.2 \textwidth, width = .95\textwidth]{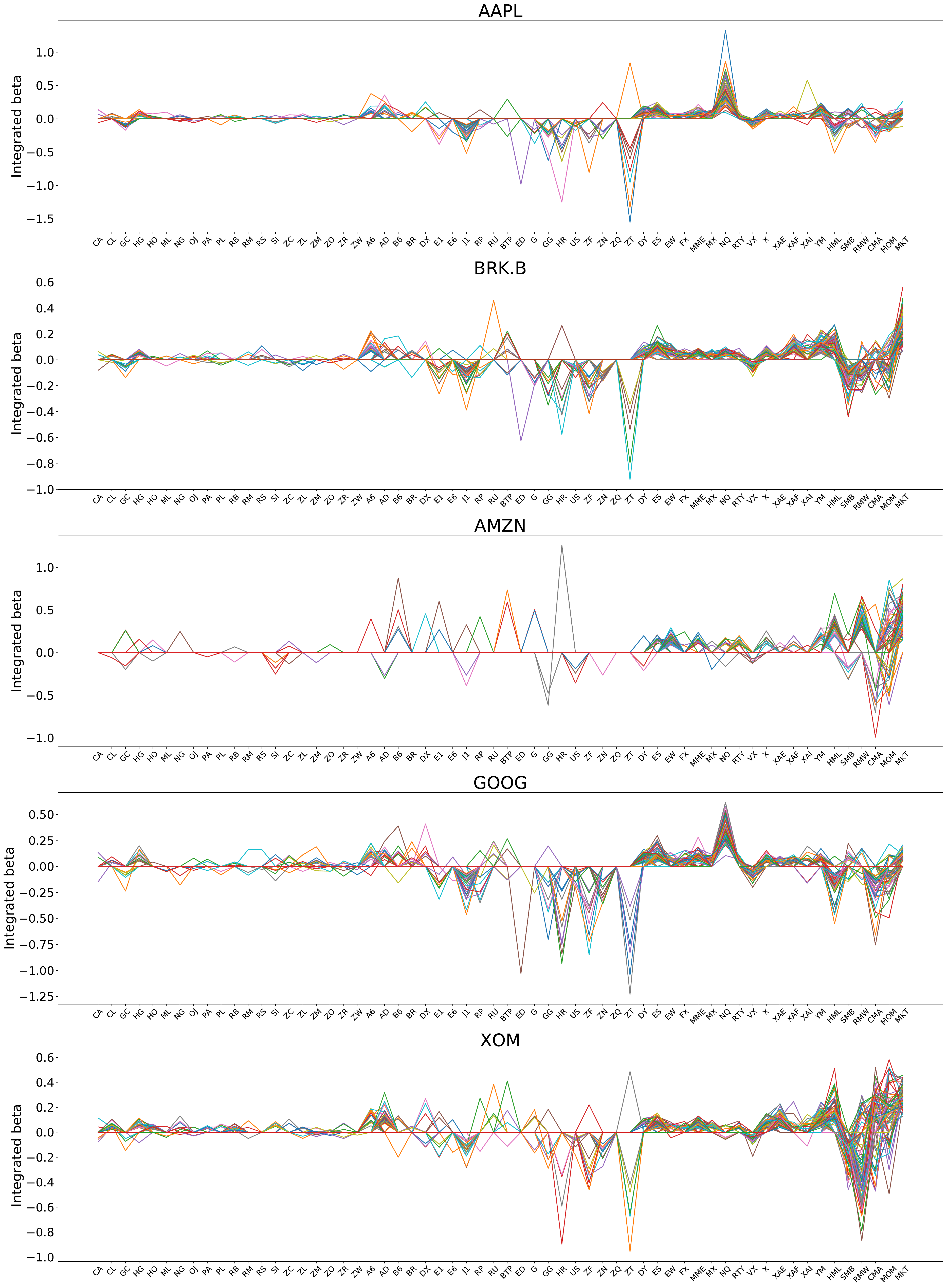}
\caption{The monthly integrated betas from the RED-LASSO estimator for the five assets and 60 factors. 
Each line connects the 60 integrated beta estimates for each month.}
\label{fig:allvalue}
\end{figure}

\begin{figure}[ht!]
\centering
\includegraphics[height=1.17 \textwidth, width = .95\textwidth]{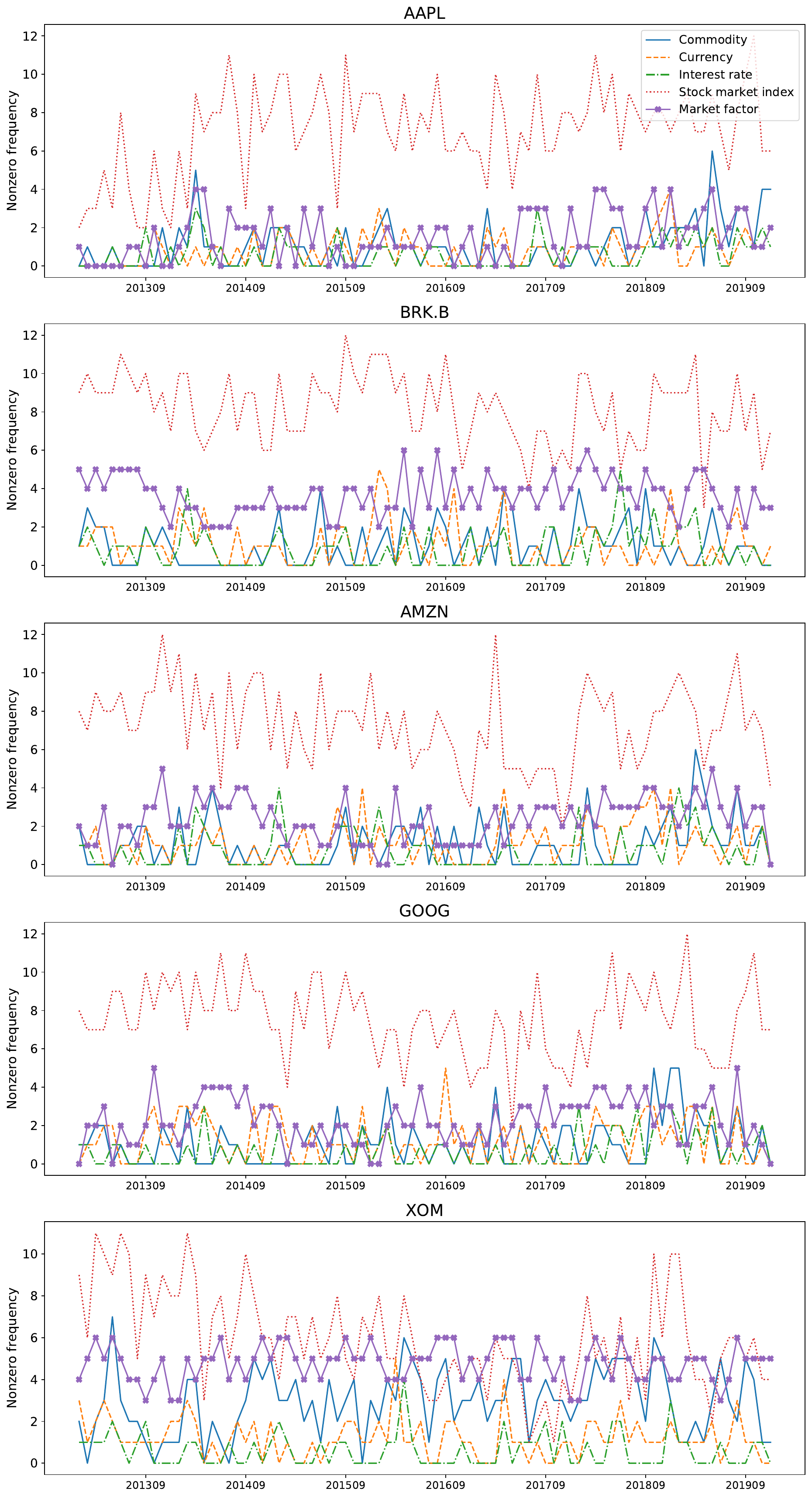}
\caption{The non-zero frequency of the monthly integrated betas from the RED-LASSO estimator for the five assets and five groups.
The five groups consist of the commodity futures group, currency futures group, interest rate futures group, stock market index futures group, and market factor group.}
\label{fig:emp_nonzero_time}
\end{figure}


\begin{figure}[!ht]
\centering
\includegraphics[height=1.2 \textwidth, width = 0.75 \textwidth]{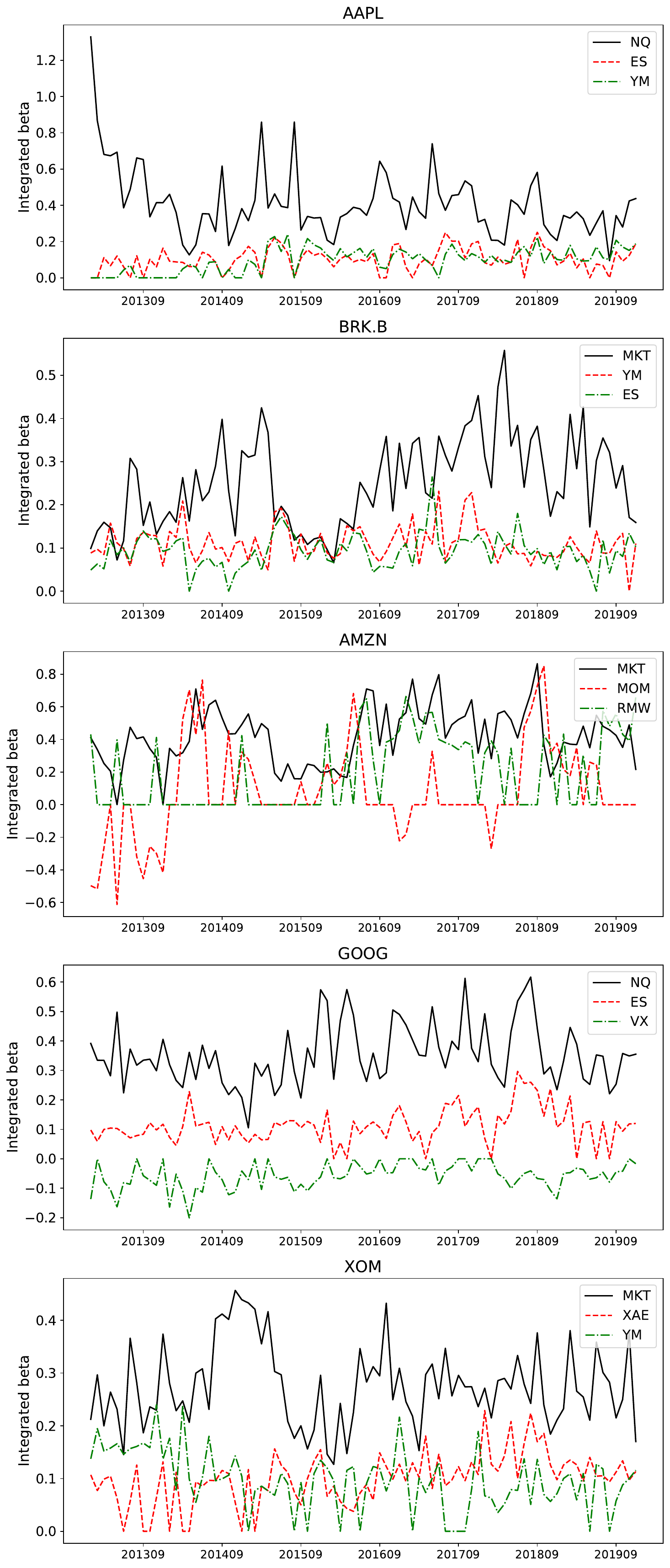}
\caption{The monthly integrated betas from the RED-LASSO estimator for the three factors that most frequently had non-zero integrated betas among the 60 factors for each of the five assets.}
\label{fig:mainfactors}
\end{figure}

Now, we investigate the result of the RED-LASSO estimator.
Figure \ref{fig:allvalue} shows the monthly integrated betas from the RED-LASSO estimator for the five assets and 60 factors. 
Figure \ref{fig:emp_nonzero_time} depicts the non-zero frequency of the RED-LASSO estimator for the five groups, consisting of the commodity futures group, currency futures group, interest rate futures group, stock market index futures group, and market factor group. 
From Figures \ref{fig:allvalue} and \ref{fig:emp_nonzero_time}, we see that integrated betas change over time, and only a small number of factors had non-zero integrated betas in most periods.
To investigate the time-series of the significant betas, we plotted the integrated beta estimates for the three factors that most frequently had non-zero integrated betas in Figure \ref{fig:mainfactors}.
The AAPL has NQ (E-mini Nasdaq 100), ES (E-mini S\&P 500), and YM (E-mini Dow); BRK.B has MKT, YM, and ES; AMZN has MKT, MOM, and RMW; GOOG has NQ, ES, and VX (VIX); and XOM has MKT, XAE (E-mini Energy Select Sector), and YM. 
In sum, either the NQ factor or MKT factor most frequently had non-zero integrated betas, while the other factors had non-zero integrated betas only for some time periods.

\begin{figure}[!ht]
\centering
\includegraphics[height=1.19 \textwidth, width = 0.97\textwidth]{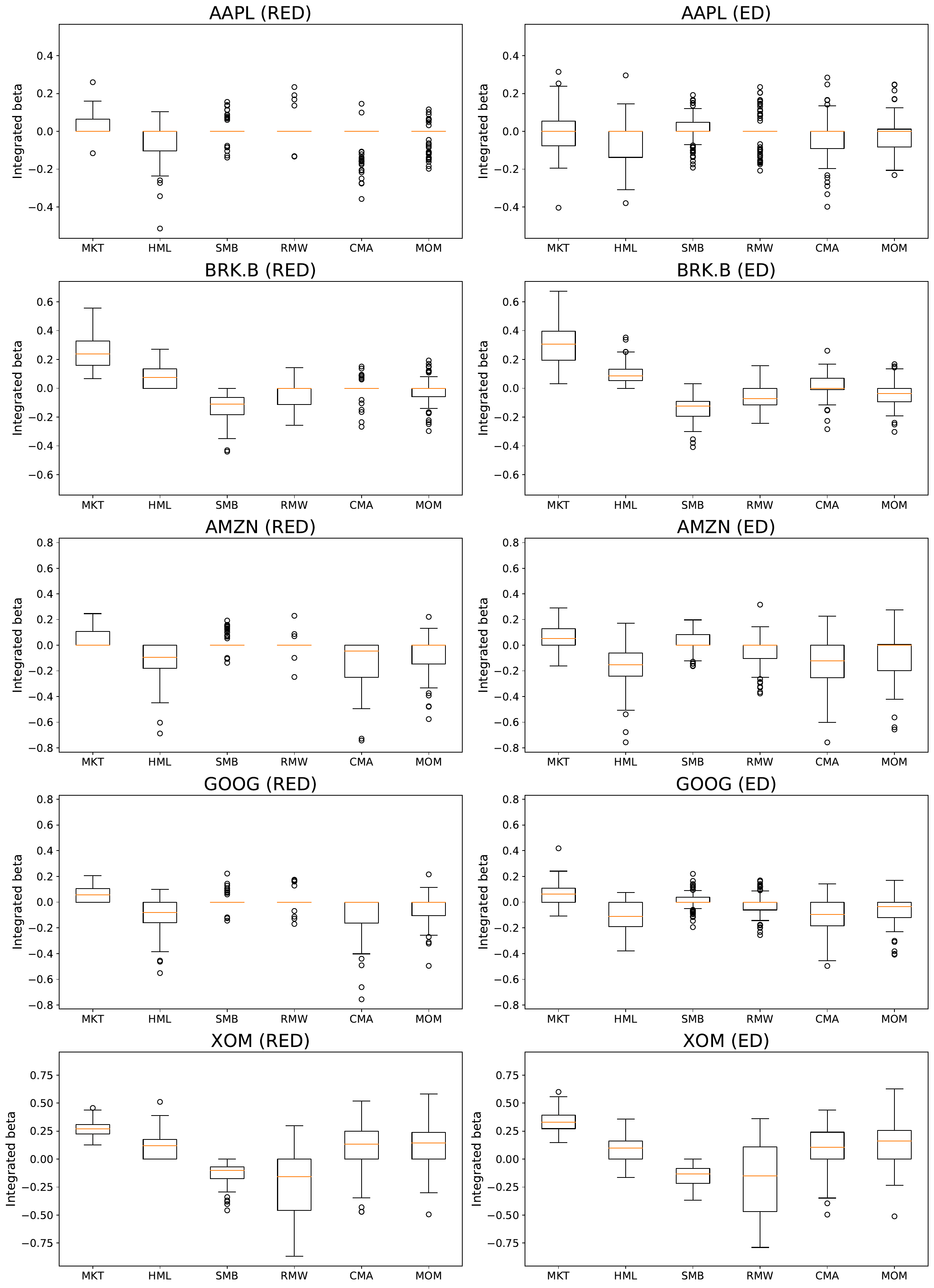}
\caption{The box plots of the monthly integrated betas from the RED-LASSO and ED-LASSO estimators for MKT, HML, SMB, RMW, CMA, and MOM over 84 months for each of the five assets.
}
\label{fig:6factors}
\end{figure}

When modeling regression-based financial models, we often employ the six factors,  MKT, HML, SMB, RMW, CMA, and MOM \citep{asness2013value, carhart1997persistence,  fama2015five, fama2016dissecting}.
To investigate their beta behaviors in more detail, we present the box plots of the integrated betas from the RED-LASSO and ED-LASSO estimators for these six factors in Figure \ref{fig:6factors}.
As expected, the MKT factor played a significant role for BRK.B and XOM. 
Specifically, the MKT factor had non-zero integrated betas for all 84 months as shown in Table \ref{Table4} in the Appendix. 
In contrast, the six factors frequently had zero integrated betas for AAPL, AMZN, and GOOG.
This may be because technology companies, such as AAPL, AMZN, and GOOG, have recently shown outstanding performance in the U.S. market, potentially reducing their dependence on these six factors.
We note that the results of the two estimators are similar, but the RED-LASSO estimator has a more stable result.
Thus, we can conjecture that considering both the heavy-tailed distribution and the time variation of the beta process helps better explain the beta dynamics.

\section{Conclusion}\label{SEC-6}
In this paper, we developed a novel RED-LASSO estimation procedure that can handle the heavy-tailedness of financial data and account for the time variation and sparsity of the high-dimensional beta process.
 To estimate the instantaneous beta, we propose a robust estimator that employs the Huber loss, truncation method, and $\ell_1$-penalty. 
 We demonstrated that the proposed instantaneous beta estimator can handle the heavy-tailedness and the curse of dimensionality with a desirable convergence rate. 
To handle the heavy-tailed bias coming from the Huber loss and  $\ell_1$-penalty, we developed a robust debiasing scheme and propose an integrated beta estimator.
We showed that the proposed debiasing method sufficiently mitigates the effect of the bias, and the integrated beta estimator can enjoy the law of large numbers property.
Then, the debiased integrated beta estimator is further regularized to account for the sparsity of the integrated beta. 
We demonstrated that the proposed RED-LASSO estimator can achieve the near-optimal convergence rate.

In the empirical study, the RED-LASSO estimation procedure shows the best performance in terms of $R^2$ and the sparsity of the beta estimates.
It suggests that when estimating integrated beta in the high-dimensional high-frequency set-up, the RED-LASSO estimation method helps account for the features of the time-varying beta process and heavy-tailed distributions of observed log-returns.
On the other hand, the proposed tuning parameter selection procedure does not guarantee the theoretical properties.
It would be interesting to develop a robust tuning parameter selection method with rigorous theoretical guarantees and practical applicability.
However, developing a procedure that satisfies both theoretical and practical criteria may be challenging.
In addition, in this paper, we did not consider microstructure noises.
The microstructure noise could be another source of the heavy tails and accommodating them leads to an application for higher frequency observations.
However, if we impose the microstructure noise structure on the regression diffusion model, we have an unbalanced order relationship between the noise and regression variables, which ruins the usual regression structure. 
Hence, it is difficult to apply the existing estimation methods.
It would be interesting and important to develop a robust estimation method that can handle microstructure noises.
Finally,  for each local sparse beta estimation, thanks to the bias adjustment, the bias is negligible, and we can obtain an asymptotic normality. 
However, the parameter of interest is the integrated beta, and the integrated beta estimator has a faster convergence rate than that of the local beta estimator, due to the property of the law of large numbers. 
Unfortunately, the convergence rate of the integrated beta estimator can be dominated by that of the bias term.
Thus, we still have a bias issue for the integrated beta inference. 
It would be interesting but difficult to develop a novel bias adjustment scheme to manage this non-negligible bias term. 
We leave these topics for future study.

\section*{Funding} 
The research of Minseok Shin was supported in part by the National Research Foundation of Korea (NRF) grant funded by the Korean government (MSIT) (RS-2025-24535699), and in part by the Institute of Information \& Communications Technology Planning \& Evaluation (IITP)-Global Data-X Leader HRD program grant funded by the Korean government (MSIT) (IITP-2025-RS-2024-00441244). 
The research of Donggyu Kim was supported in part by the National Research Foundation of Korea (NRF) grant funded by the Korean government (MSIT) (RS-2024-00343129).

\newpage
\appendix
\section{Appendix}\label{Appendix}

\subsection{Proof of Theorem \ref{Thm1}}
Without loss of generality, it is enough to show the statement for fixed $i$.
For simplicity, we denote $\bbeta_{0}(i \Delta_n)$ by $\bbeta_{0}= (\beta_{10}, \ldots, \beta_{p0})^{\top}$.

\begin{proposition}\label{Prop1}
Under the assumptions in Theorem \ref{Thm1}, we have
\begin{equation}\label{Prop1-eq1}
\left\|\nabla {\mathcal{L}}_{\tau,i}(\bbeta_{0})\right\|_{\max} \leq \eta/2,
\end{equation}
with probability greater than $1-p^{-a}$ for any given positive constant $a$. 
\end{proposition}
\textbf{Proof of Proposition \ref{Prop1}.}
Define 
\begin{equation*}
\mathcal{Y}^c_i = 
\begin{pmatrix}
\Delta_{i+1} ^n Y^c  \\ 
\Delta_{i+2} ^n Y^c  \\  
 \vdots \\ 
\Delta_{i+k_n} ^n Y^c 
\end{pmatrix}
, \quad
 \mathcal{X}_i^c = 
\begin{pmatrix}
\Delta_{i+1} ^n \bX^{c\top} \\ 
\Delta_{i+2} ^n \bX^{c\top}\\  
 \vdots \\ 
\Delta_{i+k_n} ^n \bX^{c\top}
\end{pmatrix}
, \quad \text{and} \quad
\mathcal{\tilde{X} }_i = 
\begin{pmatrix}
\int_{i \Delta_n } ^{ (i +1)  \Delta_n}   (\bbeta(t)- \bbeta_0)  ^{\top} d\bX^c(t)  \\ 
\int_{(i+1) \Delta_n } ^{(i+2)  \Delta_n}   (\bbeta(t)- \bbeta_0)  ^{\top}  d\bX^c(t) \\  
 \vdots \\ 
\int_{(i+k_n-1) \Delta_n } ^{(i+k_n)  \Delta_n}   (\bbeta(t)- \bbeta_0) ^{\top}  d\bX^c(t)
\end{pmatrix}.
\end{equation*}
We have
 \begin{eqnarray*}
 	(\mathcal{Y}^c_i)_k  &=&   \int_{(i+k-1) \Delta_n } ^{(i+k)  \Delta_n} \bbeta^{\top}(t) d\bX^c(t) + \int_{(i+k-1) \Delta_n }^{(i+k) \Delta_n}  d Z^c(t) \cr
 		&=&      \bbeta_0 ^{\top} \Delta_{i+k} ^n  \bX^c + \Delta_{i+k} ^n  Z^c  +  \int_{(i+k-1) \Delta_n } ^{(i+k)  \Delta_n}   (\bbeta(t)- \bbeta_0)^{\top} d\bX^c(t) \cr
 		&=& \left(\mathcal{X}_i^c \bbeta_0 + \mathcal{Z}_i + \mathcal{\tilde{X}}_i \right)_{k}.
 \end{eqnarray*}
Thus, for $1 \leq j \leq p$, we have 
\begin{equation}\label{Prop1-eq2}
\left|\nabla_j {\mathcal{L}}_{\tau,i}(\bbeta_{0})\right|=\left|\dfrac{\partial {\mathcal{L}}_{\tau,i}(\bbeta_{0})}{\partial \bbeta _{j}}\right| \leq (\uppercase\expandafter{\romannumeral1})_{j} +(\uppercase\expandafter{\romannumeral2})_{j},
\end{equation}
where
\begin{equation*}
(\uppercase\expandafter{\romannumeral1})_{j} = \frac{1}{k_n}\left|\sum ^{k_n}_{h=1}\psi_{\tau}\left(\mathcal{Z}_{ih} + \mathcal{\tilde{X}}_{ih} \right)\Delta_{i+h}^{n} X_j^c \right|,
\end{equation*}
\begin{eqnarray*}
(\uppercase\expandafter{\romannumeral2})_{j} &=& \frac{1}{k_n}\Biggl|\sum ^{k_n}_{h=1}\Biggl[\psi_{\tau}\left( \Delta_{i+h}^{n}Y-\langle  \Delta_{i+h}^{n} \hat{\bX}^c, \bbeta_{0} \rangle \right)\Delta_{i+h}^{n} \hat{X}_j^c  \cr
&&\qquad  \qquad \;  -\psi_{\tau}\left( \Delta_{i+h}^{n}Y^c-\langle  \Delta_{i+h}^{n} \bX^c, \bbeta_{0} \rangle \right)\Delta_{i+h}^{n} X_j^c \Biggr]\Biggr|.
\end{eqnarray*}
First, we consider $(\uppercase\expandafter{\romannumeral1})_{j}$.
By the boundedness condition Assumption \ref{assumption1}(b), we can show, with probability at least $1-p^{-2-a}$,
\begin{equation}\label{Prop1-eq3}
\sup_{1 \leq h \leq k_n} \left|\Delta_{i+h}^{n} X_j^c \right| \leq C_{X} \sqrt{ \log p} n^{-1/2}
\end{equation}
for some positive constant  $C_{X}$.
Then, we have
\begin{eqnarray}\label{Prop1-eq4}
(\uppercase\expandafter{\romannumeral1})_{j}  &\leq& \frac{1}{k_n}\sum ^{k_n}_{h=1}\left|\psi_{\tau}\left(\mathcal{Z}_{ih} + \mathcal{\tilde{X}}_{ih} \right)\1\left(|\Delta_{i+h}^{n} X_j^c | > C_X \sqrt{ \log p} n^{-1/2} \right)\right| \cr
&& + \frac{1}{k_n}\sum ^{k_n}_{h=1}\left| \mathbb{E}\left\{    \psi_{\tau}\left(\mathcal{Z}_{ih}+ \mathcal{\tilde{X}}_{ih} \right)\Delta_{i+h}^{n} X_j^c\1\left(|\Delta_{i+h}^{n} X_j^c | \leq C_X \sqrt{ \log p} n^{-1/2} \right)  \Big | \FF_{(i+h-1)\Delta_n}   \right\} \right| \cr
&& +  \frac{1}{k_n}\Big|\sum ^{k_n}_{h=1}\psi_{\tau}\left(\mathcal{Z}_{ih} + \mathcal{\tilde{X}}_{ih} \right)\Delta_{i+h}^{n} X_j^c\1\left(|\Delta_{i+h}^{n} X_j^c | \leq C_X \sqrt{ \log p} n^{-1/2} \right)\cr
&& \qquad \qquad  -\mathbb{E}\left\{    \psi_{\tau}\left(\mathcal{Z}_{ih} + \mathcal{\tilde{X}}_{ih} \right)\Delta_{i+h}^{n} X_j^c\1\left(|\Delta_{i+h}^{n} X_j^c | \leq C_X \sqrt{ \log p} n^{-1/2} \right)  \Big | \FF_{(i+h-1)\Delta_n}   \right\} \Big| \cr
&=& (\uppercase\expandafter{\romannumeral1})_{j}^{(1)} + (\uppercase\expandafter{\romannumeral1})_{j}^{(2)} + (\uppercase\expandafter{\romannumeral1})_{j}^{(3)}.
\end{eqnarray}
For $(\uppercase\expandafter{\romannumeral1})_{j}^{(1)}$, by \eqref{Prop1-eq3}, we have
\begin{equation}\label{Prop1-eq5}
\Pr \left\{ (\uppercase\expandafter{\romannumeral1})_{j}^{(1)} = 0 \right\} \geq 1-p^{-2-a}.
\end{equation}
For $(\uppercase\expandafter{\romannumeral1})_{j}^{(2)}$, let $f(h)=\psi_{\tau}\left(\mathcal{Z}_{ih}+ \mathcal{\tilde{X}}_{ih} \right)\Delta_{i+h}^{n} X_j^c\1\left(|\Delta_{i+h}^{n} X_j^c | \leq C_X \sqrt{ \log p} n^{-1/2} \right)$. Then, we have
\begin{eqnarray*}
&&\left| \mathbb{E}\left\{    f(h)  \Big | \FF_{(i+h-1)\Delta_n}   \right\}\right| \cr
&& \leq \left| \mathbb{E}\left\{  \left(\mathcal{Z}_{ih}+ \mathcal{\tilde{X}}_{ih} \right)\Delta_{i+h}^{n} X_j^c\1\left(|\Delta_{i+h}^{n} X_j^c | \leq C_X \sqrt{ \log p} n^{-1/2} \right)  \Big | \FF_{(i+h-1)\Delta_n}   \right\}\right| \cr
&& \quad +  \left| \mathbb{E}\left\{ f(h)  - \left(\mathcal{Z}_{ih} + \mathcal{\tilde{X}}_{ih} \right)  \Delta_{i+h}^{n} X_j^c\1\left(|\Delta_{i+h}^{n} X_j^c | \leq C_X \sqrt{ \log p} n^{-1/2} \right)  \Big | \FF_{(i+h-1)\Delta_n}   \right\}\right|.
\end{eqnarray*}
Consider the first term. Similar to the proofs of Theorem 1 \citep{Kim2024regression}, we can show, for any constant $b \geq 1$, 
\begin{eqnarray}\label{Prop1-eq6}
&& \Pr \left\{ \sup_{1 \leq h \leq k_n} \mathbb{E}\left\{|\mathcal{\tilde{X}}_{ih}|^{b}  \Big | \FF_{(i+h-1)\Delta_n} \right\} \leq  \(C s_p \Delta_n \sqrt{b k_n \log p}\)^{b}\right\} \geq 1-p^{-2-a} \quad \text{ and } \cr
&&  \sup_{1 \leq h \leq k_n}\mathbb{E}\left\{|\Delta_{i+h}^{n} X_j^c |^{b} \Big | \FF_{(i+h-1)\Delta_n} \right\} \leq  \left(C \Delta_n b\right)^{b/2} \text{ a.s.}
\end{eqnarray}
Then, by Cauchy–Schwarz inequality, we have, with probability at least $1-p^{-2-a}$,
\begin{eqnarray*}
&& \sup_{1 \leq h \leq k_n}\left|\mathbb{E} \left\{\mathcal{\tilde{X}}_{ih} \Delta_{i+h}^{n} X_j^c\1\left(|\Delta_{i+h}^{n} X_j^c | \leq C_X \sqrt{ \log p} n^{-1/2} \right) \Big | \FF_{(i+h-1)\Delta_n} \right\} \right| \cr
&& \leq \sup_{1 \leq h \leq k_n} \sqrt{\mathbb{E}\left\{ | \mathcal{\tilde{X}}_{ih}|^2 \Big | \FF_{(i+h-1)\Delta_n} \right\}}\sup_{1 \leq h \leq k_n} \sqrt{\mathbb{E}\left\{ |\Delta_{i+h}^{n} X_j^c|^2 \Big | \FF_{(i+h-1)\Delta_n} \right\}} \cr 
&& \leq C s_p \Delta_n^{3/2}\sqrt{k_n \log p}.
\end{eqnarray*}
Also, for $1 \leq h \leq k_n$, we have
  \begin{eqnarray*}
\mathbb{E}\left\{\mathcal{Z}_{ih} \Delta_{i+h}^{n} X_j^c \1\left(|\Delta_{i+h}^{n} X_j^c | \leq C_X \sqrt{ \log p} n^{-1/2} \right) \Big | \FF_{(i+h-1)\Delta_n}  \right\} = 0.
  \end{eqnarray*}
Thus, we have, with probability at least $1-p^{-2-a}$,
  \begin{eqnarray}\label{Prop1-eq7}
&& \sup_{1 \leq h \leq k_n}\left| \mathbb{E}\left\{  \left(\mathcal{Z}_{ih}+ \mathcal{\tilde{X}}_{ih} \right)\Delta_{i+h}^{n} X_j^c\1\left(|\Delta_{i+h}^{n} X_j^c | \leq C_X \sqrt{ \log p} n^{-1/2} \right)  \Big | \FF_{(i+h-1)\Delta_n}   \right\} \right| \cr 
&& \leq C s_p \Delta_n^{3/2}\sqrt{k_n \log p}.
  \end{eqnarray}
Consider the second term. 
By \eqref{Prop1-eq6} and H\"older's inequality, we have, with probability at least $1-p^{-2-a}$,
\begin{equation}\label{Prop1-eq8}
\sup_{1 \leq h \leq k_n}\mathbb{E}\left\{| \mathcal{Z}_{ih} + \mathcal{\tilde{X}}_{ih} |^{\gamma} \Big | \FF_{(i+h-1)\Delta_n} \right\} \leq C n^{-\gamma/2}.
\end{equation}
Then, using the fact that
\begin{eqnarray*}
 \left| \psi_{\tau}\left(\mathcal{Z}_{ih} + \mathcal{\tilde{X}}_{ih} \right) - \left(\mathcal{Z}_{ih} + \mathcal{\tilde{X}}_{ih} \right)  \right| &\leq& \left| \mathcal{Z}_{ih} + \mathcal{\tilde{X}}_{ih} \right| \1\left( \left| \mathcal{Z}_{ih} + \mathcal{\tilde{X}}_{ih} \right| > \tau \right) \cr
 &\leq&  \tau^{-1}  \left| \mathcal{Z}_{ih} + \mathcal{\tilde{X}}_{ih} \right|^2 \text{ a.s.},
\end{eqnarray*}
we have,  with probability at least $1-p^{-2-a}$,
\begin{eqnarray}\label{Prop1-eq9}
  && \sup_{1 \leq h \leq k_n} \left| \mathbb{E}\left\{  f(h)  - \left(\mathcal{Z}_{ih} + \mathcal{\tilde{X}}_{ih} \right)  \Delta_{i+h}^{n} X_j^c\1\left(|\Delta_{i+h}^{n} X_j^c | \leq C_X \sqrt{ \log p} n^{-1/2} \right)  \Big | \FF_{(i+h-1)\Delta_n}   \right\} \right| \cr
  && \leq \tau^{-1}  \sup_{1 \leq h \leq k_n}\mathbb{E} \left\{  | \mathcal{Z}_{ih} + \mathcal{\tilde{X}}_{ih} |^2 |\Delta_{i+h}^{n} X_j^c  | \Big | \FF_{(i+h-1)\Delta_n} \right\}  \cr
  && \leq \tau^{-1} \sup_{1 \leq h \leq k_n} \left( \mathbb{E} \left\{ | \mathcal{Z}_{ih} + \mathcal{\tilde{X}}_{ih} |^{\gamma} \Big | \FF_{(i+h-1)\Delta_n} \right\} \right)^{2/\gamma} \left( \mathbb{E} \left\{|\Delta_{i+h}^{n} X_j^c  |^{\gamma / (\gamma-2)} \Big | \FF_{(i+h-1)\Delta_n} \right\} \right)^{(\gamma-2)/ \gamma} \cr
   && \leq C\tau^{-1} n^{-3/2},
\end{eqnarray}
where the second inequality is due to  H\"older's inequality and the last inequality is from \eqref{Prop1-eq6} and \eqref{Prop1-eq8}.
By \eqref{Prop1-eq7} and \eqref{Prop1-eq9}, we have
\begin{equation}\label{Prop1-eq10}
\Pr\left\{(\uppercase\expandafter{\romannumeral1})_{j}^{(2)} \leq C \left(s_p \Delta_n^{3/2}\sqrt{k_n \log p} +  \tau^{-1} n^{-3/2} \right)\right\} \geq 1-2p^{-2-a}.
\end{equation}
For $(\uppercase\expandafter{\romannumeral1})_{j}^{(3)}$, by (2.1) in \citet{freedman1975tail}, we have
\begin{eqnarray}\label{Prop1-eq11}
&& \Pr\left\{(\uppercase\expandafter{\romannumeral1})_{j}^{(3)} \geq s  \text{ and }  \sum ^{k_n}_{h=1} \mathbb{E}\left\{ (f(h))^2 \Big | \FF_{(i+h-1)\Delta_n}\right\} \leq C n^{-2} k_n  \right\} \cr
&& \leq 2 \exp\left\{-Ck_{n}^{2}s^2 \Big/ \(\tau \sqrt{\log p}n^{-1/2}k_n s + n^{-2}k_n\)\right\}.
\end{eqnarray}
Also, by \eqref{Prop1-eq6} and \eqref{Prop1-eq8}, we have, with probability at least $1-p^{-2-a}$,
\begin{eqnarray*}
  && \sup_{1 \leq h \leq k_n} \left| \mathbb{E}\left\{ (f(h))^2 \Big | \FF_{(i+h-1)\Delta_n}   \right\} \right| \cr
  && \leq \sup_{1 \leq h \leq k_n} \mathbb{E} \left\{  | \mathcal{Z}_{ih} + \mathcal{\tilde{X}}_{ih} |^2 |\Delta_{i+h}^{n} X_j^c  |^2 \Big | \FF_{(i+h-1)\Delta_n} \right\}  \cr
   && \leq  \left(  \sup_{1 \leq h \leq k_n} \mathbb{E} \left\{ | \mathcal{Z}_{ih} + \mathcal{\tilde{X}}_{ih} |^{\gamma} \Big | \FF_{(i+h-1)\Delta_n} \right\} \right)^{2/\gamma} \left( \sup_{1 \leq h \leq k_n}  \mathbb{E} \left\{|\Delta_{i+h}^{n} X_j^c  |^{2\gamma / (\gamma-2)} \Big | \FF_{(i+h-1)\Delta_n} \right\} \right)^{(\gamma-2)/ \gamma} \cr
   && \leq C n^{-2},
\end{eqnarray*}
where the second inequality is due to H\"older's inequality.
Thus, we have
\begin{equation}\label{Prop1-eq12}
Pr\left\{(\uppercase\expandafter{\romannumeral1})_{j}^{(3)} \leq C\left(\tau n^{-1/2}k_n^{-1} (\log p)^{3/2} + n^{-1}k_n^{-1/2}\sqrt{\log p} \right)\right\} \geq  1-2p^{-2-a}.
\end{equation}
By \eqref{Prop1-eq4}, \eqref{Prop1-eq5}, \eqref{Prop1-eq10}, and \eqref{Prop1-eq12}, we have, with probability at least $1-5p^{-1-a}$,
\begin{equation}\label{Prop1-eq13}
\max_{1 \leq j \leq p}(\uppercase\expandafter{\romannumeral1})_{j} \leq C \left[s_p \Delta_n^{3/2}\sqrt{k_n \log p} +  \tau^{-1} n^{-3/2}+\tau n^{-1/2}k_n^{-1} (\log p)^{3/2}\right].
\end{equation}

Consider $(\uppercase\expandafter{\romannumeral2})_{j}$.
For some large constant $C>0$, define 
\begin{eqnarray*}
&& Q_1=\{ \max_{i, j}  | \Delta_{i} ^n X_j^c |\leq   C \sqrt{ \log p} n^{-1/2} \}, \cr
&& Q_2=\{ \max_{i,j}  \int _{i \Delta_n} ^{\(i+k_n\) \Delta_n} d \Lambda_{j}(t)  \leq  C  \log p \} \cap  \{ \max_{i}  \int _{i \Delta_n} ^{\(i+k_n\) \Delta_n} d \Lambda^y(t)  \leq  C  \log p \}, \cr
&& Q_3=\{ \max_{i,j}\sum_{ k=1}^ {k_n} \1 _{\{ | \Delta_{i+k} ^n X_j | > v_{j,n}\} }  \leq C \log p \}.
\end{eqnarray*}
By \eqref{Prop1-eq3}, we have
\begin{equation*}
\Pr\( Q_1 \) \geq 1- p^{-2-a}.
\end{equation*}
By the boundedness of the intensity process, we have
\begin{equation*}
\Pr\( Q_2 \) \geq 1- p^{-2-a}.
\end{equation*}
Under the event $Q_1 \cap Q_2$, we have, for large $n$,
\begin{eqnarray*}
\max_{i,j}\sum_{ k=1}^ {k_n} \1 _{\{ | \Delta_{i+k} ^n X_j | > v_{j,n}\} } \leq \max_{i,j}  \int _{i \Delta_n} ^{\(i+k_n\) \Delta_n} d \Lambda_{j}(t) \leq C \log p.
\end{eqnarray*}
Thus, we have 
\begin{equation} \label{Prop1-eq14}
	\Pr \( Q_1 \cap Q_2 \cap Q_3\) \geq 1 - 2p^{-2-a}. 
\end{equation}
We note that, for any $x_{1}, x_{2}, y_{1}, y_{2} \in \mathbb{R}$,
\begin{equation*}
\left|x_{1}y_{1}-x_{2}y_{2}\right| \leq \left|\(x_{1}-x_{2}\)\(y_{1}-y_{2}\)\right|+ \left|\(x_{1}-x_{2}\)y_{2}\right| +  \left|x_{2}\(y_{1}-y_{2}\)\right|.
\end{equation*}
Hence, under the event $ Q_1 \cap Q_2 \cap Q_3$, we have 
\begin{eqnarray*}
&&(\uppercase\expandafter{\romannumeral2})_{j} \cr
&&\leq \frac{1}{k_n}\sum ^{k_n}_{h=1}\Biggl[\left|\psi_{\tau}\left(\Delta_{i+h}^{n}Y-\langle  \Delta_{i+h}^{n} \hat{\bX}^c, \bbeta_{0} \rangle \right) -\psi_{\tau}\left(\Delta_{i+h}^{n}Y^c-\langle  \Delta_{i+h}^{n} \bX^c, \bbeta_{0} \rangle\right)\right|  \left| \Delta_{i+h}^{n} \hat{X}_j^c - \Delta_{i+h}^{n} X_j^c \right|\cr
&&  \quad  \qquad \qquad  + \left|\psi_{\tau}\left(\Delta_{i+h}^{n}Y-\langle  \Delta_{i+h}^{n} \hat{\bX}^c, \bbeta_{0} \rangle \right) -\psi_{\tau}\left(\Delta_{i+h}^{n}Y^c-\langle  \Delta_{i+h}^{n} \bX^c, \bbeta_{0} \rangle\right)\right|  \left| \Delta_{i+h}^{n} X_j^c \right|  \cr
&& \qquad  \qquad \quad +  \left|\psi_{\tau}\left( \Delta_{i+h}^{n}Y^c-\langle  \Delta_{i+h}^{n} \bX^c, \bbeta_{0} \rangle \right) \right|  \left| \Delta_{i+h}^{n} \hat{X}_j^c - \Delta_{i+h}^{n} X_j^c \right| \Biggr] \cr
&&   \leq \frac{C}{k_n} \sum ^{k_n}_{h=1}\Biggl[\tau  \left| \Delta_{i+h}^{n} \hat{X}_j^c - \Delta_{i+h}^{n} X_j^c \right| \cr
&& \qquad \qquad \quad + \sqrt{\log p} n^{-1/2}  \left|\psi_{\tau}\left(\Delta_{i+h}^{n}Y-\langle  \Delta_{i+h}^{n} \hat{\bX}^c, \bbeta_{0} \rangle \right) -\psi_{\tau}\left(\Delta_{i+h}^{n}Y^c-\langle  \Delta_{i+h}^{n} \bX^c, \bbeta_{0} \rangle\right)\right| \Biggr] \cr
&&\leq C \left\{\tau n^{-1}(\log p)^{3/2} + \tau n^{-1} (\log p )^{3/2} + k_n^{-1}\sqrt{\log p} n^{-1/2} \sum ^{k_n}_{h=1} \left|\langle  \Delta_{i+h}^{n} \hat{\bX}^c-  \Delta_{i+h} \bX^c, \bbeta_{0} \rangle \right| \right\} \cr
&& \leq  C \left\{\tau n^{-1}(\log p)^{3/2}  + s_p n^{-3/2}(\log p)^2 \right\}\text{ a.s.},
\end{eqnarray*}
which implies
\begin{equation}\label{Prop1-eq15}
\Pr \(\max_{1 \leq j \leq p}(\uppercase\expandafter{\romannumeral2})_{j} \leq C \left\{\tau n^{-1}(\log p)^{3/2}  + s_p n^{-3/2}(\log p)^2\right\} \) \geq 1-2p^{-1-a}.
\end{equation}
Combining \eqref{Prop1-eq2}, \eqref{Prop1-eq13}, and \eqref{Prop1-eq15}, we have,  with probability greater than $1-p^{-a}$,
\begin{equation}\label{Prop1-eq16}
\left\|\nabla {\mathcal{L}}_{\tau,i}(\bbeta_{0})\right\|_{\infty} \leq C \left[s_p \Delta_n^{3/2}\sqrt{k_n \log p} +  \tau^{-1} n^{-3/2}+\tau n^{-1/2}k_n^{-1} (\log p)^{3/2}\right].
\end{equation}
\endpf

\textbf{Proof of Theorem \ref{Thm1}.} 
By Proposition \ref{Prop1}, it is enough to show the statement under \eqref{Prop1-eq1}.
First, we investigate  $\hat{\bbeta}_{i \Delta_n} - \bbeta_{0}$. Since 
\begin{equation*}
{\mathcal{L}}_{\tau,i}(\hat{\bbeta}_{i \Delta_n}) + \eta \|\hat{\bbeta}_{i \Delta_n} \| _{1} \leq  {\mathcal{L}}_{\tau,i}(\bbeta_{0}) + \eta \left\|\bbeta_{0} \right\| _{1},
\end{equation*}
we have
\begin{eqnarray*}
\eta\left( \left\|\bbeta_{0} \right\| _{1} - \|\hat{\bbeta}_{i \Delta_n} \| _{1}\right) &\geq& {\mathcal{L}}_{\tau,i}(\hat{\bbeta}_{i \Delta_n})  - {\mathcal{L}}_{\tau,i}(\bbeta_{0}) \cr
&\geq& \langle \nabla {\mathcal{L}}_{\tau,i}(\bbeta_{0}), \, \hat{\bbeta}_{i \Delta_n} - \bbeta_{0}  \rangle \cr
&\geq& -\eta \|\hat{\bbeta}_{i \Delta_n} - \bbeta_{0} \|_{1}/2.
\end{eqnarray*}
Then, we have
\begin{eqnarray*}
&& \|(\hat{\bbeta}_{i \Delta_n}- \bbeta_{0})_{S_{i\Delta_n}} \| _{1} + \|(\hat{\bbeta}_{i \Delta_n}- \bbeta_{0})_{S_{i\Delta_n}^c} \| _{1} \cr 
&& = \|\hat{\bbeta}_{i \Delta_n} - \bbeta_{0} \|_{1} \cr
&& \geq 2 \left(\|\hat{\bbeta}_{i \Delta_n} \|_{1}  -\| \bbeta_{0} \|_{1}  \right) \cr
&& = 2 \left(\|(\hat{\bbeta}_{i \Delta_n})_{S_{i\Delta_n}^c} \|_{1} + \|(\hat{\bbeta}_{i \Delta_n})_{S_{i\Delta_n}} \|_{1} - \|({\bbeta}_{0})_{S_{i\Delta_n}} \|_{1} - \|({\bbeta}_{0})_{S_{i\Delta_n}^c} \|_{1}\right)   \cr
&& \geq 2\left(\|(\hat{\bbeta}_{i \Delta_n}- \bbeta_{0})_{S_{i\Delta_n}^c} \|_{1} - \|(\hat{\bbeta}_{i \Delta_n}- \bbeta_{0})_{S_{i\Delta_n}} \|_{1} -2 \|(\bbeta_{0})_{S_{i\Delta_n}^c} \|_{1}\right).
\end{eqnarray*}
Thus, we have
\begin{equation}\label{Thm1-eq1}
\hat{\bbeta}_{i \Delta_n} - \bbeta_{0} \in  \mathcal W_{i \Delta_n}, 
\end{equation}
where $W_{i \Delta_n}$ is defined in Assumption \ref{assumption1}(e).

Now, we investigate $\| \hat{\bbeta}_{i \Delta_n} - \bbeta_{0}\|_{1}$ and $\| \hat{\bbeta}_{i \Delta_n} - \bbeta_{0}\|_{2}$.
By \eqref{sparsity_beta}, we have
\begin{equation}\label{Thm1-eq2}
(n\eta)^{\delta}|S_{i\Delta_n}| \leq s_p \quad \text{ and } \quad \|(\bbeta_{0})_{S_{i\Delta_n}^c} \|_{1} \leq \Sigma_{j \in S_{i\Delta_n}^c}|(\bbeta_{0})_j|^{\delta} |(\bbeta_{0})_j|^{1-\delta} \leq s_p(n\eta)^{1-\delta}. 
\end{equation}
Thus, by \eqref{Thm1-eq1}--\eqref{Thm1-eq2}, we have
\begin{eqnarray}\label{Thm1-eq3}
\| \hat{\bbeta}_{i \Delta_n} - \bbeta_{0}\|_{1} &\leq&  4 \| (\hat{\bbeta}_{i \Delta_n} - \bbeta_{0})_{S_{i\Delta_n}}\|_{1} + 4\| (\bbeta_{0})_{S_{i\Delta_n}^c}\|_{1}\cr
&\leq&  4\sqrt{s_p}(n\eta)^{-\delta/2}\| \hat{\bbeta}_{i \Delta_n} - \bbeta_{0}\|_{2} + 4s_p (n\eta)^{1-\delta},
\end{eqnarray}
where the second inequality is due to Cauchy–Schwarz inequality. Suppose that
\begin{equation}\label{Thm1-eq4}
\| \hat{\bbeta}_{i \Delta_n} - \bbeta_{0}\|_{2} > (1+12/ \kappa)\sqrt{s_p}(n \eta)^{1-\delta/2}.
\end{equation}
Then, we have
\begin{equation}\label{Thm1-eq5}
\| \hat{\bbeta}_{i \Delta_n} - \bbeta_{0}\|_{1} < \frac{8\kappa+48}{\kappa+12} \sqrt{s_p}(n \eta)^{-\delta/2} \| \hat{\bbeta}_{i \Delta_n} - \bbeta_{0}\|_{2}.
\end{equation}
From the optimality of $\hat{\bbeta}_{i \Delta_n}$ and the integral form of the Taylor expansion, we have
\begin{eqnarray}\label{Thm1-eq6}
0 &\geq& {\mathcal{L}}_{\tau,i}(\hat{\bbeta}_{i \Delta_n})  - {\mathcal{L}}_{\tau,i}(\bbeta_{0}) + \eta \left(  \|\hat{\bbeta}_{i \Delta_n} \| _{1}  - \left\|\bbeta_{0} \right\| _{1} \right) \cr
&=& \eta \left(  \|\hat{\bbeta}_{i \Delta_n} \| _{1}  - \left\|\bbeta_{0} \right\| _{1} \right) + \langle  \nabla {\mathcal{L}}_{\tau,i}(\bbeta_{0}), \, \hat{\bbeta}_{i \Delta_n} - \bbeta_{0} \rangle \cr
&& + \int^{1}_{0} \left(1-t\right)(\hat{\bbeta}_{i \Delta_n} - \bbeta_{0})^{\top}\nabla^{2}{\mathcal{L}}_{\tau,i}(\bbeta_{0}+t(\hat{\bbeta}_{i \Delta_n} - \bbeta_{0}))(\hat{\bbeta}_{i \Delta_n} - \bbeta_{0}) dt.
\end{eqnarray}
For the first and second terms, we have
\begin{eqnarray}\label{Thm1-eq7}
&& \eta \left(  \|\hat{\bbeta}_{i \Delta_n} \| _{1}  - \left\|\bbeta_{0} \right\| _{1} \right) + \langle  \nabla {\mathcal{L}}_{\tau,i}(\bbeta_{0}), \, \hat{\bbeta}_{i \Delta_n} - \bbeta_{0} \rangle \cr
&& \geq -\eta \|\hat{\bbeta}_{i \Delta_n} -\bbeta_{0} \| _{1} -\|\nabla {\mathcal{L}}_{\tau,i}(\bbeta_{0}) \|_{\max}  \|\hat{\bbeta}_{i \Delta_n} -\bbeta_{0} \| _{1} \cr
&& \geq -\frac{12\kappa +72}{\kappa+12}\sqrt{s_p}n^{-\delta/2}\eta^{1-\delta/2}\|\hat{\bbeta}_{i \Delta_n}  -\bbeta_{0} \| _{2}, 
\end{eqnarray}
where the second inequality is due to \eqref{Thm1-eq5}.
For the last term, let
\begin{equation*}
z= \frac{(\kappa+12)(n \eta)^{\delta/2}D}{(8\kappa+48)\sqrt{s_p}\|\hat{\bbeta}_{i \Delta_n}  -\bbeta_{0} \| _{2}} < 1.
\end{equation*}
Then, for any $0 \leq t \leq z$, we have 
\begin{equation*}
\|\bbeta_{0} + t(\hat{\bbeta}_{i \Delta_n}-\bbeta_{0}) -\bbeta_{0} \| _{1} \leq z \|\hat{\bbeta}_{i \Delta_n} -\bbeta_{0} \| _{1} \leq D,
\end{equation*}
where the last inequality is due to \eqref{Thm1-eq5}.
Thus, by Assumption \ref{assumption1}(e), we have
\begin{eqnarray}\label{Thm1-eq8}
&& \int^{1}_{0} \left(1-t\right)(\hat{\bbeta}_{i \Delta_n} - \bbeta_{0})^{\top}\nabla^{2}{\mathcal{L}}_{\tau,i}(\bbeta_{0}+t(\hat{\bbeta}_{i \Delta_n} - \bbeta_{0}))(\hat{\bbeta}_{i \Delta_n} - \bbeta_{0}) dt \cr
&& \geq  \int^{z}_{0} (1-t)\kappa n^{-1} \|\hat{\bbeta}_{i \Delta_n}  -\bbeta_{0} \| _{2}^2 dt  \cr
&& = (\kappa+12)\sqrt{s_p}n^{-\delta/2}\eta^{1-\delta/2}\|\hat{\bbeta}_{i \Delta_n}  -\bbeta_{0} \|_{2} - \frac{(\kappa+12)^2}{2 \kappa}s_p  n^{1-\delta} \eta^{2-\delta}.
\end{eqnarray}
Combining \eqref{Thm1-eq6}--\eqref{Thm1-eq8}, we have
\begin{equation*}
\frac{(\kappa+12)^2}{2 \kappa}s_p n^{1-\delta}\eta^{2-\delta} \geq \frac{\kappa^2 + 12 \kappa + 72}{\kappa+12}\sqrt{s_p}n^{-\delta/2}\eta^{1-\delta/2}\|\hat{\bbeta}_{i \Delta_n}  -\bbeta_{0} \| _{2},
\end{equation*}
which implies
\begin{eqnarray*}
&& \|\hat{\bbeta}_{i \Delta_n}  -\bbeta_{0} \| _{2} \leq \frac{(\kappa+12)^2}{2(\kappa^2 + 12 \kappa + 72)}\frac{\kappa+12}{\kappa}\sqrt{s_p}(n\eta)^{1-\delta/2} \cr
&& \leq (1+12/ \kappa)\sqrt{s_p}(n\eta)^{1-\delta/2}.
\end{eqnarray*}
This contradicts to \eqref{Thm1-eq4}, thus, we obtain the $\ell_2$ norm error bound. Then, by \eqref{Thm1-eq3}, we can show the $\ell_1$ norm error bound.
\endpf

\subsection{Proof of Theorem \ref{Thm2}}
\textbf{Proof of Theorem \ref{Thm2}.} 
We first investigate $\hat{\bbeta}_{i \Delta_n}$ and $\hat{\bOmega}_{i \Delta_n}$. 
By \eqref{sparsity_beta}, \eqref{Thm1-result1}, and Assumption \ref{assumption1}(c), we can show, with probability at least $1-p^{-2-a}$,
\begin{eqnarray}\label{Thm2-eq1}
&& \sup_{0 \leq i \leq n-k_n} \| \hat{\bbeta}_{i \Delta_n} - \bbeta_{0}((i+k_n) \Delta_n) \|_{1} \leq Cs_p\left(s_p n^{-1/4}\(\log p\)^{3/4} \right)^{1-\delta} \quad \text{ and }\cr
&&  \sup_{0 \leq i \leq n-k_n} \| \hat{\bbeta}_{i \Delta_n} \|_{1} \leq Cs_p.
\end{eqnarray}
For $\hat{\bOmega}_{i \Delta_n}$, similar to the proofs of Theorem 1 \citep{Kim2024regression}, we can show, with probability at least $1-p^{-2-a}$,
\begin{equation}\label{Thm2-eq2}
\sup_{0 \leq i \leq n-k_n} \|   \frac{1}{k_n \Delta_n }    \mathcal{X}_{i}  ^{\top}\mathcal{X}_{i} \bOmega_0(i\Delta_n) - \bI  \|_{\max} \leq \lambda.
\end{equation}
Thus, we have, with probability at least $1-p^{-2-a}$,
\begin{equation}\label{Thm2-eq3}
\sup_{0 \leq i \leq n-k_n} \| \hat{\bOmega}_{i \Delta_n} \|_{1} \leq C. 
\end{equation}
Consider $\tilde{\bbeta}_{i \Delta_n}$.
For each $1 \leq m \leq p$, there exists standard Brownian motion $W_{m}^*(t)$  such that 
  \begin{equation*}
 d \beta_{m}(t) = \mu_{\beta,m}(t) dt + \sqrt{ \Sigma_{\beta,mm}(t)} dW_{m}^{*}(t).
  \end{equation*}
Then, by the proofs of Theorem 1 \citep{Kim2024regression}, we have
\begin{equation*}
\frac{1}{k_n \Delta_n}\mathcal{X}_i^{c\top}\mathcal{\tilde{X}}_{i} = \frac{1}{k_n \Delta_n}\mathcal{A}_i  + R_i,
\end{equation*}
where
\begin{eqnarray*}
&& \mathcal{A}_i = \(  \sum_{m=1}^p    \int_{ i \Delta_n } ^{ \(i+k_n\) \Delta_n}   \int_{i \Delta_n} ^{t} \sqrt{\Sigma_{\beta,mm}(s)} dW_{m}^{*}(s)    \Sigma_{jm}(t)  d t \)_{j=1, \ldots,p} \quad \text{ and} \cr
&& \Pr\left\{\sup_{0 \leq i \leq n-k_n} \| R_i \|_{\max}  \leq  C s_p n^{-1/2}\(\log p\)^{3/2} \right\} \geq 1-p^{-2-a}.
\end{eqnarray*}
Note that
\begin{equation*}
\Pr \left\{\sup_{0 \leq i \leq n-k_n} \left\|\tilde{\mathcal{X}_i}\right\|_{\max} \leq   C s_p n^{-3/4} \log p  \right\} \geq 1-p^{-2-a}.
\end{equation*}
Hence, similar to the proofs of \eqref{Prop1-eq15}, we can show
\begin{equation*}
\frac{1}{k_n \Delta_n}\mathcal{X}_i^{\top}\mathcal{\tilde{X}}_{i} = \frac{1}{k_n \Delta_n}\mathcal{A}_i  + R_i^{'},
\end{equation*}
where
\begin{equation}\label{Thm2-eq4}
\Pr\left\{\sup_{0 \leq i \leq n-k_n} \| R_i^{'} \|_{\max}  \leq  C s_p n^{-1/2}\(\log p\)^{3/2} \right\} \geq 1-p^{-1-a}.
\end{equation}

Let
\begin{eqnarray*}
&& \tilde{ \bbeta}_{i \Delta_n}^{(2)} = \hat{\bbeta}_{i \Delta_n}  +  \psi_{\varpi}\left(\frac{1}{k_n \Delta_n }   \hat{\bOmega}_{i \Delta_n}^{\top} \mathcal{X}_{(i+k_n)}  ^{\top}  \left( \mathcal{Y}_{(i+k_n)}^{c} - \mathcal{X}_{(i+k_n)}  \hat{\bbeta}_{i \Delta_n} \right)\right), \cr
&& \tilde{ \bbeta}_{i \Delta_n}^{(3)} =\hat{\bbeta}_{i \Delta_n}  +  \psi_{\varpi}\(\frac{1}{k_n \Delta_n }   \hat{\bOmega}_{i \Delta_n}^{\top} \mathcal{X}_{(i+k_n)}^{\top}  \Big[\mathcal{X}_{(i+k_n)}^{c}\bbeta_{0}((i+k_n) \Delta_n)  - \mathcal{X}_{(i+k_n)} \hat{\bbeta}_{i \Delta_n}\Big] +\mathcal{B}_{i+k_n}   \),  \cr
&& \tilde{ \bbeta}_{i \Delta_n}^{(4)} =\hat{\bbeta}_{i \Delta_n}  +  \psi_{\varpi}\(\frac{1}{k_n \Delta_n }   \hat{\bOmega}_{i \Delta_n}^{\top} \mathcal{X}_{(i+k_n)}^{c\top} \mathcal{X}_{(i+k_n)}^{c} \Big[\bbeta_{0}((i+k_n) \Delta_n)  -  \hat{\bbeta}_{i \Delta_n}\Big] +\mathcal{B}_{i+k_n}   \), \cr
&& \tilde{ \bbeta}_{i \Delta_n}^{(5)} =\hat{\bbeta}_{i \Delta_n}  +  \psi_{\varpi}\(\bbeta_{0}((i+k_n) \Delta_n)  -  \hat{\bbeta}_{i \Delta_n} +\mathcal{B}_{i+k_n}   \),
\end{eqnarray*}
where
\begin{equation*}
\mathcal{B}_{i}=\frac{1}{k_n \Delta_n }   \hat{\bOmega}_{(i-k_n) \Delta_n}^{\top} \left(\mathcal{X}_{i}^{\top}\mathcal{Z}_i + \mathcal{A}_i  \right).
\end{equation*}
Then, we have
\begin{eqnarray}\label{Thm2-eq5}
\|  \hat{I\beta}  - I\beta_0 \| _{\max} &\leq&   \left\|  \sum_{i=0}^{[1/(k_n \Delta_n) ]-2}\left(\tilde{\bbeta}_{i k_n \Delta_n} - \tilde{\bbeta}_{i k_n \Delta_n}^{(2)} \right) k_n \Delta_n \right\|_{\max}    \cr
&&+  \left\|  \sum_{i=0}^{[1/(k_n \Delta_n) ]-2}\left(\tilde{\bbeta}_{i k_n \Delta_n}^{(2)} - \tilde{\bbeta}_{i k_n \Delta_n}^{(3)} \right) k_n \Delta_n \right\|_{\max}     \cr
&&+  \left\|  \sum_{i=0}^{[1/(k_n \Delta_n) ]-2}\left(\tilde{\bbeta}_{i k_n \Delta_n}^{(3)} - \tilde{\bbeta}_{i k_n \Delta_n}^{(4)} \right) k_n \Delta_n \right\|_{\max}     \cr
&&+  \left\|  \sum_{i=0}^{[1/(k_n \Delta_n) ]-2}\left(\tilde{\bbeta}_{i k_n \Delta_n}^{(4)} - \tilde{\bbeta}_{i k_n \Delta_n}^{(5)} \right) k_n \Delta_n \right\|_{\max}     \cr
&&+ \left\|  \sum_{i=0}^{[1/(k_n \Delta_n) ]-2}\left(\tilde{\bbeta}_{i k_n \Delta_n}^{(5)} - \bbeta_0((i+1)k_n\Delta_n) \right) k_n \Delta_n  \right\|_{\max}     \cr
&&+   \left\|  \sum_{i=0}^{[1/(k_n \Delta_n) ]-2} \int_{i k_n \Delta_n}^{(i+1) k_n \Delta_n}\left(\bbeta_0((i+1)k_n\Delta_n) -\bbeta_0(t) \right)  dt \right\|_{\max}    \cr
&&+   \left\|  \int_{([1/(k_n \Delta_n) ]-1)k_n\Delta_n}^{1} \bbeta_0(t) dt \right\|_{\max}    \cr
& = & (\uppercase\expandafter{\romannumeral1})+(\uppercase\expandafter{\romannumeral2})+(\uppercase\expandafter{\romannumeral3})+(\uppercase\expandafter{\romannumeral4})+(\uppercase\expandafter{\romannumeral5})+(\uppercase\expandafter{\romannumeral6})+(\uppercase\expandafter{\romannumeral7}).
\end{eqnarray}
Consider $(\uppercase\expandafter{\romannumeral1})$. By the boundedness of the intensity, we can show $\Pr\left\{ \int _{0} ^{1} d \Lambda^y(t)  \leq  C \log p \right\} \geq 1-p^{-1-a}$. Thus, we have
\begin{equation}\label{Thm2-eq6}
\Pr\left\{(\uppercase\expandafter{\romannumeral1}) \leq Cs_p^{2-\delta}n^{(-2+\delta)/4}(\log p)^{(5-3\delta)/4}\right\} \geq 1-p^{-1-a}.
\end{equation}
For $(\uppercase\expandafter{\romannumeral2})$, by \eqref{Thm2-eq3}--\eqref{Thm2-eq4}, we have, with probability at least $1-2p^{-1-a}$,
\begin{eqnarray}\label{Thm2-eq7}
(\uppercase\expandafter{\romannumeral2}) &\leq& \sup_{0 \leq i \leq n-2k_n}\left\|\hat{\bOmega}_{i \Delta_n}^{\top} \left(\frac{1}{k_n \Delta_n}\mathcal{X}_{i+k_n}^{\top}\mathcal{\tilde{X}}_{i+k_n} - \frac{1}{k_n \Delta_n}\mathcal{A}_{i+k_n}  \right)\right\|_{\max} \cr
&\leq& C s_p n^{-1/2}\(\log p\)^{3/2}.
\end{eqnarray}
Consider $(\uppercase\expandafter{\romannumeral3})$. Similar to the proofs of (A.20) in \citet{Kim2024regression}, we can show,  with probability at least $1-p^{-1-a}$,
\begin{eqnarray}\label{Thm2-eq8}
(\uppercase\expandafter{\romannumeral3}) &\leq&  \sum_{i=0}^{[1/(k_n \Delta_n) ]-2} \left\|   \hat{\bOmega}_{i \Delta_n}^{\top} \left( \mathcal{X}_{(i+k_n)}^{\top}  \mathcal{X}_{(i+k_n)}^{c} - \mathcal{X}_{(i+k_n)}^{c\top}  \mathcal{X}_{(i+k_n)}^{c}\right) \bbeta_{0}((i+k_n) \Delta_n)  \right\|_{\max}   \cr
&&+ \sum_{i=0}^{[1/(k_n \Delta_n) ]-2} \left\|\hat{\bOmega}_{i \Delta_n}^{\top} \left(\mathcal{X}_{(i+k_n)}^{\top}  \mathcal{X}_{(i+k_n)} - \mathcal{X}_{(i+k_n)}^{c\top}  \mathcal{X}_{(i+k_n)}^{c}\right)\hat{\bbeta}_{i \Delta_n}  \right\|_{\max} \cr
&\leq& C s_p n^{-3/4}\sqrt{\log p}.
\end{eqnarray}
Consider $(\uppercase\expandafter{\romannumeral4})$. By Assumption \ref{assumption1}(b) and (f), we can show, with probability at least $1-p^{-1-a}$,
\begin{eqnarray*}
&&\sup_{0 \leq i \leq n-k_n}\left\|\bSigma_0(i\Delta_n) - \frac{1}{k_n \Delta_n }\mathcal{X}_{i}^{c\top}  \mathcal{X}_{i}^{c} \right\|_{\max} \cr
&& \leq \sup_{0 \leq i \leq n-k_n}\left\|\bSigma_0(i\Delta_n) - \frac{1}{k_n \Delta_n }\int_{i\Delta_n}^{(i+k_n)\Delta_n}\bSigma_0(t)dt \right\|_{\max} \cr
&& \quad + \sup_{0 \leq i \leq n-k_n}\left\|\frac{1}{k_n \Delta_n }\int_{i\Delta_n}^{(i+k_n)\Delta_n}\bSigma_0(t)dt - \frac{1}{k_n \Delta_n }\mathcal{X}_{i}^{c\top}  \mathcal{X}_{i}^{c} \right\|_{\max} \cr
&&\leq Cn^{-1/4}\sqrt{\log p}.
\end{eqnarray*}
Thus, by Assumption \ref{assumption1}(f), we can show,  with probability at least $1-p^{-1-a}$,
\begin{equation*}
\sup_{0 \leq i \leq n-k_n}\left\|\frac{1}{k_n \Delta_n }\left(\mathcal{X}_{(i+k_n)}^{c\top}  \mathcal{X}_{(i+k_n)}^{c} - \mathcal{X}_{i}^{c\top}  \mathcal{X}_{i}^{c} \right) \right\|_{\max} \leq Cn^{-1/4}\sqrt{\log p}.
\end{equation*}
Then, by \eqref{Prop1-eq14}, \eqref{Thm2-eq1}, and \eqref{Thm2-eq3}, we have,  with probability at least $1-p^{-1-a}$,
\begin{eqnarray}\label{Thm2-eq9}
(\uppercase\expandafter{\romannumeral4}) &\leq&  \sup_{0 \leq i \leq n-2k_n}\left\|\frac{1}{k_n \Delta_n } \hat{\bOmega}_{i \Delta_n}^{\top} \mathcal{X}_{(i+k_n)}^{c\top} \mathcal{X}_{(i+k_n)}^{c}  -\bI \right\|_{\max}  \times \sup_{0 \leq i \leq n-2k_n}\left\| \bbeta_{0}((i+k_n) \Delta_n)  -  \hat{\bbeta}_{i \Delta_n} \right\|_{1}  \cr
&\leq&  \sup_{0 \leq i \leq n-2k_n} \Bigg[\left\|\frac{1}{k_n \Delta_n } \hat{\bOmega}_{i \Delta_n}^{\top} \left(\mathcal{X}_{(i+k_n)}^{c\top}  \mathcal{X}_{(i+k_n)}^{c} - \mathcal{X}_{i}^{c\top}  \mathcal{X}_{i}^{c}  \right) \right\|_{\max} \cr
&& \qquad \quad \qquad + \left\|\frac{1}{k_n \Delta_n } \hat{\bOmega}_{i \Delta_n}^{\top} \left(\mathcal{X}_{i}^{c\top}  \mathcal{X}_{i}^{c} -\mathcal{X}_{i}^{\top}  \mathcal{X}_{i} \right) \right\|_{\max} +\left\| \frac{1}{k_n \Delta_n } \hat{\bOmega}_{i \Delta_n}^{\top} \mathcal{X}_{i}^{\top}  \mathcal{X}_{i} -\bI \right\|_{\max} \Bigg] \cr
&& \times Cs_p\left(s_p n^{-1/4}\(\log p\)^{3/4} \right)^{1-\delta}\cr
&\leq& C \left(n^{-1/4} \sqrt{\log p} + n^{-1/2}(\log p)^2 + \lambda    \right) \times s_p\left(s_p n^{-1/4}\(\log p\)^{3/4} \right)^{1-\delta} \cr
&\leq& Cs_p^{2-\delta}n^{(-2+\delta)/4}(\log p)^{(5-3\delta)/4}.
\end{eqnarray}
For $(\uppercase\expandafter{\romannumeral5})$, let $g(i)=\bbeta_{0}((i+k_n) \Delta_n)  -  \hat{\bbeta}_{i \Delta_n} +\mathcal{B}_{i+k_n}$. We have
\begin{eqnarray*}
(\uppercase\expandafter{\romannumeral5}) &\leq&  \left\|  \sum_{i=0}^{[1/(k_n \Delta_n) ]-2}\left[\psi_{\varpi}\left(g(ik_n)\right) - \mathbb{E}\left\{\psi_{\varpi}\left(g(ik_n)\right) \Big | \FF_{(i+1)k_n\Delta_n} \right\} \right] k_n \Delta_n  \right\|_{\max} \cr
&& + \left\|  \sum_{i=0}^{[1/(k_n \Delta_n) ]-2}\left[\mathbb{E}\left\{\psi_{\varpi}\left(g(ik_n)\right) \Big | \FF_{(i+1)k_n\Delta_n} \right\} - \bbeta_{0}((i+1)k_n \Delta_n)  +  \hat{\bbeta}_{ik_n \Delta_n}  \right] k_n \Delta_n  \right\|_{\max} \cr
&& = \left\|(\uppercase\expandafter{\romannumeral5})^{(1)}\right\|_{\max}+\left\|(\uppercase\expandafter{\romannumeral5})^{(2)}\right\|_{\max}.
\end{eqnarray*}
For the first term, by the boundedness of the intensity process and \eqref{Thm2-eq3}, we can show, with probability at least  $1-p^{-2-a}$,
\begin{equation*}
\sup_{k_n \leq i \leq n-k_n}\sup_{1 \leq j \leq p}\mathbb{E}\left\{\mathcal{B}_{ij}^2 \Big | \FF_{i\Delta_n} \right\} \leq Cs_{p}^{2}n^{-1/2}.
\end{equation*}
Thus, from \eqref{Thm2-eq1}, we have, with probability at least  $1-2p^{-2-a}$,
\begin{equation*}
\sup_{0 \leq i \leq n-2k_n}\sup_{1 \leq j \leq p}\mathbb{E}\Big[ |(g(i))_j|^2 \Big | \FF_{(i+k_n)\Delta_n} \Big]    \leq Cs_p^{2}\left(s_p n^{-1/4}\(\log p\)^{3/4} \right)^{2-2\delta}.
\end{equation*}
Then, by (2.1) in \citet{freedman1975tail}, we have, for $1 \leq j \leq p$,
\begin{eqnarray*}
&& \Pr\Bigg[(\uppercase\expandafter{\romannumeral5})_{j}^{(1)} \geq s  \, \text{ and }  \sum_{i=0}^{[1/(k_n \Delta_n) ]-2} \mathbb{E}\Big[|\left(g(ik_n)\right)_j|^2 \Big | \FF_{(i+1)k_n\Delta_n}\Big] \leq Cs_p^{2}n^{1/2}\left(s_p n^{-1/4}\(\log p\)^{3/4} \right)^{2-2\delta}  \Bigg] \cr
&& \leq 2 \exp\left\{-Cns^2 \Big/ \(n^{1/2}\varpi s + s_p^{4-2\delta}n^{\delta/2}\(\log p\)^{(3-3\delta)/2}  \)\right\},
\end{eqnarray*}
which implies
\begin{equation*}
\Pr \Bigg[\left\|(\uppercase\expandafter{\romannumeral5})^{(1)}\right\|_{\max} \leq Cs_p^{2-\delta}n^{(-2+\delta)/4}(\log p)^{(5-3\delta)/4} \Bigg] \geq 1-3p^{-1-a}.
\end{equation*}
For the second term, we have, with probability at least  $1-2p^{-2-a}$,
\begin{eqnarray*}
&& \left\|(\uppercase\expandafter{\romannumeral5})^{(2)}\right\|_{\max} \cr
&& \leq \sup_{0 \leq i \leq [1/(k_n \Delta_n) ]-2}\sup_{1 \leq j \leq p} \left| \mathbb{E}\left\{\left[\psi_{\varpi}\left(g(ik_n)\right)\right]_j \Big | \FF_{(i+1)k_n\Delta_n} \right\} - \left[\bbeta_{0}((i+1)k_n \Delta_n)  - \hat{\bbeta}_{ik_n \Delta_n}  \right]_j \right| \cr
&& = \sup_{0 \leq i \leq [1/(k_n \Delta_n) ]-2}\sup_{1 \leq j \leq p} \left|\mathbb{E}\left\{\left[\psi_{\varpi}\left(g(ik_n)\right)-g(ik_n)\right]_j \Big | \FF_{(i+1)k_n\Delta_n} \right\}  \right| \cr
&& \leq \sup_{0 \leq i \leq [1/(k_n \Delta_n) ]-2}\sup_{1 \leq j \leq p} \mathbb{E}\left\{\left|\left[g(ik_n)\right]_j\right|\1\(\left|\left[g(ik_n)\right]_j\right|>\varpi\) \Big | \FF_{(i+1)k_n\Delta_n} \right\}  \cr
&& \leq \sup_{0 \leq i \leq [1/(k_n \Delta_n) ]-2}\sup_{1 \leq j \leq p} \mathbb{E}\left\{\left|\left[g(ik_n)\right]_j\right|^2 /\varpi \Big | \FF_{(i+1)k_n\Delta_n} \right\}   \cr
&&\leq Cs_p^{2-\delta}n^{(-2+\delta)/4}(\log p)^{(5-3\delta)/4}.
\end{eqnarray*}
Thus, we have
\begin{equation}\label{Thm2-eq10}
\Pr\left\{(\uppercase\expandafter{\romannumeral5}) \leq Cs_p^{2-\delta}n^{(-2+\delta)/4}(\log p)^{(5-3\delta)/4} \right\} \geq 1-4p^{-1-a}.
\end{equation}
Consider $(\uppercase\expandafter{\romannumeral6})$. By the sub-Gaussianity of the beta process, we can show, with probability at least $1-p^{-1-a}$,
\begin{equation}\label{Thm2-eq11}
 (\uppercase\expandafter{\romannumeral6}) \leq C\sqrt{\log p /n}. 
\end{equation}
For $(\uppercase\expandafter{\romannumeral7})$, by Assumption \ref{assumption1}(b), we have
\begin{equation}\label{Thm2-eq12}
 (\uppercase\expandafter{\romannumeral7}) \leq Cn^{-1/2} \text{ a.s.} 
\end{equation}
Combining \eqref{Thm2-eq5}--\eqref{Thm2-eq12}, we have, with probability greater than $1-p^{-a}$,
\begin{equation}\label{Thm2-eq13}
 \|  \hat{I\beta}  - I\beta_0 \| _{\max} \leq C\left[s_p^{2-\delta}n^{(-2+\delta)/4}(\log p)^{(5-3\delta)/4} + s_p n^{-1/2}\(\log p\)^{3/2}\right].
\end{equation}
\endpf

\subsection{Proof of Theorem \ref{Thm3}}
\textbf{Proof of Theorem \ref{Thm3}.}
By \eqref{Thm2-result1}, there exists $h_n$ such that, with probability greater than $1-p^{-a}$,
\begin{equation*}
 \|  \hat{I\beta}  - I\beta_0 \| _{\max}  \leq h_n/2.
\end{equation*}
Thus, it is enough to show the statement under the event $\{\|  \hat{I\beta}  - I\beta_0 \| _{\max}  \leq h_n/2 \}$.
Similar to the proofs of Theorem 1 \citep{Kim2024regression}, we can show
\begin{eqnarray*} 
	\|  \tilde{I\beta}  - I\beta_0 \|_1 \leq  C s_p h_n^{1-\delta}.  
\end{eqnarray*}
\endpf

\begin{table}[!ht]
\scalebox{0.77}{
\begin{tabular}{lllll}
\hline
\multicolumn{1}{l}{Type} & & \multicolumn{1}{l}{Symbol} &  & \multicolumn{1}{l}{Description} \\ \hline  
Commodity  
&  &  CA  &  & Cocoa                                 \\  
&  &  CL  &  & Crude Oil WTI                         \\    
&  &  GC  &  & Gold                                  \\    
&  &  HG  &  & Copper                                \\  
&  &  HO  &  & NY Harbor ULSD (Heating Oil)          \\ 
&  &  ML  &  & Milling Wheat                         \\      
&  &  NG  &  & Henry Hub Natural Gas                 \\
&  &  OJ  &  & Orange Juice                          \\  
&  &  PA  &  & Palladium                             \\  
&  &  PL  &  & Platinum                              \\
&  &  RB  &  & RBOB Gasoline                         \\
&  &  RM  &  & Robusta Coffee                        \\  
&  &  RS  &  & Canola                                \\  
&  &  SI  &  & Silver                                \\
&  &  ZC  &  & Corn                                  \\  
&  &  ZL  &  & Soybean Oil                           \\  
&  &  ZM  &  & Soybean Meal                          \\  
&  &  ZO  &  & Oats                                  \\  
&  &  ZR  &  & Rough Rice                            \\  
&  &  ZW  &  & Wheat                                 \\ 

&  &      &  &                                       \\
                     
Currency  
&  &  A6  &  & Australian Dollar                     \\
&  &  AD  &  & Canadian Dollar                       \\       
&  &  B6  &  & British Pound                         \\         
&  &  BR  &  & Brazilian Real                        \\
&  &  DX  &  & US Dollar Index                       \\        
&  &  E1  &  & Swiss Franc                           \\          
&  &  E6  &  & Euro FX                               \\          
&  &  J1  &  & Japanese Yen                          \\
&  &  RP  &  & Euro/British Pound                    \\      
&  &  RU  &  & Russian Ruble                         \\  

&  &      &  &                                       \\       
         
Interest rate  
&  &  BTP &  & Euro BTP Long-Bond                    \\  
&  &  ED  &  & Eurodollar                            \\   
&  &  G   &  & 10-Year Long Gilt                     \\    
&  &  GG  &  & Euro Bund                             \\   
&  &  HR  &  & Euro Bobl                             \\
&  &  US  &  & 30-Year US Treasury Bond              \\
&  &  ZF  &  & 5-Year US Treasury Note               \\
&  &  ZN  &  & 10-Year US Treasury Note              \\
&  &  ZQ  &  & 30-Day Fed Funds                      \\
&  &  ZT  &  & 2-Year US Treasury Note               \\   

&  &      &  &                                       \\  
                                                                             
Stock market index  
&  &  DY  &  & DAX                                   \\  
&  &  ES  &  & E-mini S\&P 500                       \\     
&  &  EW  &  & E-mini S\&P 500 Midcap                \\                            
&  &  FX  &  & Euro Stoxx 50                         \\                                                                                                                                                   
&  &  MME &  & MSCI Emerging Markets Index           \\ 
&  &  MX  &  & CAC 40                                \\          
&  &  NQ  &  & E-mini Nasdaq 100                     \\
&  &  RTY &  & E-mini Russell 2000                   \\  
&  &  VX  &  & VIX                                   \\
&  &  X   &  & FTSE 100                              \\
&  &  XAE &  & E-mini Energy Select Sector           \\
&  &  XAF &  & E-mini Financial Select Sector        \\
&  &  XAI &  & E-mini Industrial Select Sector       \\
&  &  YM  &  & E-mini Dow                            \\  \hline
\end{tabular}
\caption{The symbols of 54 futures in Section \ref{SEC-5}.}
\label{Table3}
}
\end{table}

\begin{table}[!ht]
\scalebox{0.49}{
\begin{tabular}{llll cccccccccccccccccccccccc}
\hline
\multicolumn{1}{l}{Type} && \multicolumn{1}{l}{Symbol} && \multicolumn{4}{c}{AAPL} && \multicolumn{4}{c}{BRK.B} && \multicolumn{4}{c}{AMZN} && \multicolumn{4}{c}{GOOG} && \multicolumn{4}{c}{XOM} \\   \cline{5-8} \cline{10-13} \cline{15-18} \cline{20-23} \cline{25-28}  &&  && \text{RED} & \text{ED} & \text{LASSO} & \text{SVR} && \text{RED} & \text{ED} & \text{LASSO} & \text{SVR} && \text{RED} & \text{ED} & \text{LASSO} & \text{SVR} && \text{RED} & \text{ED} & \text{LASSO} & \text{SVR} && \text{RED} & \text{ED} & \text{LASSO} & \text{SVR} \\ \hline
Commodity 
&& CA && 4 & 28 & 9  & 84 && 3  & 34 & 21 & 84 && 2 & 23 & 15 & 84 && 3 & 21 & 15 & 84 && 6 & 32 & 33 & 84 \\
&& CL && 5 & 36 & 13 & 84 && 6  & 27 & 38 & 84 && 7 & 35 & 27 & 84 && 6 & 39 & 21 & 84 && 64& 65 & 79 & 84 \\
&& GC && 15& 36 & 16 & 84 && 11 & 36 & 40 & 84 && 8 & 41 & 32 & 84 && 6 & 42 & 21 & 84 && 3 & 41 & 32 & 84 \\
&& HG && 21& 36 & 27 & 84 && 17 & 21 & 46 & 84 && 13& 38 & 30 & 84 && 13& 31 & 29 & 84 && 25& 46 & 57 & 84 \\
&& HO && 4 & 33 & 11 & 84 && 2  & 32 & 34 & 84 && 3 & 27 & 24 & 84 && 1 & 30 & 20 & 84 && 38& 47 & 76 & 84 \\
&& ML && 1 & 19 & 4  & 49 && 3  & 17 & 10 & 49 && 0 & 14 & 11 & 49 && 3 & 22 & 11 & 49 && 3 & 11 & 16 & 49 \\
&& NG && 6 & 37 & 9  & 84 && 3  & 28 & 23 & 84 && 4 & 24 & 17 & 84 && 2 & 18 & 11 & 84 && 9 & 29 & 35 & 84 \\
&& OJ && 2 & 23 & 9  & 84 && 4  & 26 & 21 & 84 && 7 & 22 & 16 & 84 && 3 & 26 & 12 & 84 && 4 & 30 & 29 & 84 \\
&& PA && 1 & 26 & 11 & 84 && 6  & 30 & 28 & 84 && 4 & 32 & 19 & 84 && 5 & 26 & 14 & 84 && 8 & 28 & 38 & 84 \\
&& PL && 5 & 28 & 8  & 84 && 5  & 27 & 23 & 84 && 7 & 30 & 12 & 84 && 2 & 28 & 15 & 84 && 5 & 30 & 41 & 84 \\
&& RB && 7 & 35 & 13 & 84 && 1  & 28 & 37 & 84 && 2 & 38 & 24 & 84 && 5 & 32 & 19 & 84 && 44& 40 & 73 & 84 \\
&& RM && 0 & 15 & 7  & 52 && 3  & 11 & 15 & 52 && 4 & 11 & 13 & 52 && 3 & 14 & 12 & 52 && 3 & 21 & 20 & 52 \\
&& RS && 2 & 25 & 9  & 84 && 5  & 24 & 25 & 84 && 3 & 19 & 14 & 84 && 2 & 15 & 13 & 84 && 0 & 17 & 25 & 84 \\
&& SI && 4 & 27 & 13 & 84 && 3  & 26 & 30 & 84 && 4 & 33 & 14 & 84 && 6 & 26 & 16 & 84 && 6 & 31 & 38 & 84 \\
&& ZC && 2 & 28 & 10 & 84 && 5  & 24 & 17 & 84 && 2 & 34 & 17 & 84 && 6 & 35 & 20 & 84 && 3 & 32 & 36 & 84 \\
&& ZL && 3 & 21 & 8  & 84 && 4  & 27 & 22 & 84 && 4 & 32 & 17 & 84 && 6 & 27 & 14 & 84 && 9 & 33 & 34 & 84 \\
&& ZM && 3 & 24 & 8  & 84 && 2  & 36 & 22 & 84 && 2 & 33 & 17 & 84 && 9 & 30 & 15 & 84 && 6 & 36 & 30 & 84 \\
&& ZO && 3 & 27 & 11 & 84 && 1  & 37 & 27 & 84 && 4 & 36 & 14 & 84 && 5 & 29 & 13 & 84 && 4 & 34 & 28 & 84 \\
&& ZR && 3 & 36 & 12 & 84 && 4  & 23 & 22 & 84 && 5 & 23 & 15 & 84 && 5 & 40 & 13 & 84 && 5 & 25 & 29 & 84 \\
&& ZW && 3 & 30 & 7  & 84 && 0  & 26 & 21 & 84 && 6 & 41 & 19 & 84 && 3 & 29 & 17 & 84 && 1 & 30 & 33 & 84 \\

&&   &&    &    &   &   &&    &    &   &   &&    &    &   &   &&    &    &   &    &&    &    &    &                        \\

Currency 
&& A6 && 10 & 40 & 23 & 84 && 17 & 43 & 46 & 84 && 14 & 30 & 29 & 84 && 17 & 37 & 26 & 84 && 18 & 44 & 59 & 84 \\
&& AD && 14 & 38 & 23 & 84 && 12 & 32 & 41 & 84 && 13 & 37 & 33 & 84 && 13 & 38 & 27 & 84 && 32 & 34 & 68 & 84 \\
&& B6 && 3  & 37 & 12 & 84 && 4  & 29 & 32 & 84 && 1  & 38 & 22 & 84 && 6  & 31 & 16 & 84 && 4  & 38 & 38 & 84 \\
&& BR && 5  & 23 & 13 & 83 && 5  & 34 & 23 & 83 && 4  & 27 & 18 & 83 && 8  & 31 & 15 & 83 && 3  & 31 & 35 & 83 \\ 
&& DX && 3  & 42 & 10 & 84 && 2  & 31 & 32 & 84 && 4  & 38 & 20 & 84 && 5  & 32 & 20 & 84 && 6  & 35 & 35 & 84 \\
&& E1 && 5  & 29 & 13 & 84 && 12 & 42 & 36 & 84 && 6  & 33 & 25 & 84 && 7  & 36 & 22 & 84 && 6  & 38 & 33 & 84 \\
&& E6 && 1  & 33 & 12 & 84 && 3  & 39 & 31 & 84 && 4  & 36 & 22 & 84 && 2  & 32 & 18 & 84 && 2  & 37 & 38 & 84 \\
&& J1 && 24 & 41 & 38 & 84 && 26 & 40 & 58 & 84 && 40 & 56 & 44 & 84 && 35 & 46 & 34 & 84 && 25 & 36 & 49 & 84 \\
&& RP && 4  & 30 & 8  & 84 && 7  & 25 & 27 & 84 && 5  & 24 & 23 & 84 && 8  & 21 & 13 & 84 && 2  & 32 & 34 & 84 \\
&& RU && 1  & 18 & 7  & 67 && 2  & 23 & 14 & 67 && 6  & 30 & 19 & 67 && 5  & 31 & 10 & 67 && 3  & 25 & 30 & 67 \\

&&   &&    &    &   &  &&    &    &   &   &&    &    &  &    &&    &    &    &    &&    &    &    &     \\  

Interest rate 
&& BTP && 2 & 24 & 5  & 84 && 8  & 30 & 27 & 84 && 8  & 34 & 16 & 84 && 4 & 27 & 12 & 84 && 4 & 29 & 31 & 84 \\
&& ED  && 1 & 2  & 2  & 36 && 1  & 2  & 10 & 36 && 2  & 3  & 8  & 36 && 1 & 3  & 7  & 36 && 0 & 3  & 11 & 36 \\
&& G   && 4 & 34 & 14 & 84 && 7  & 26 & 35 & 84 && 7  & 35 & 14 & 84 && 1 & 41 & 17 & 84 && 5 & 29 & 35 & 84 \\
&& GG  && 12& 31 & 17 & 84 && 7  & 36 & 42 & 84 && 5  & 31 & 27 & 84 && 9 & 34 & 20 & 84 && 5 & 36 & 45 & 84 \\
&& HR  && 8 & 29 & 16 & 84 && 8  & 28 & 31 & 84 && 6  & 27 & 25 & 84 && 9 & 29 & 18 & 84 && 4 & 31 & 36 & 84 \\
&& US  && 7 & 36 & 23 & 84 && 11 & 34 & 51 & 84 && 10 & 31 & 39 & 84 && 11& 33 & 26 & 84 && 7 & 32 & 36 & 84 \\
&& ZF  && 5 & 33 & 20 & 84 && 15 & 33 & 48 & 84 && 6  & 33 & 31 & 84 && 10& 36 & 28 & 84 && 9 & 31 & 45 & 84 \\
&& ZN  && 6 & 37 & 20 & 84 && 12 & 34 & 49 & 84 && 11 & 31 & 28 & 84 && 9 & 36 & 30 & 84 && 11& 31 & 42 & 84 \\
&& ZQ  && 0 & 0  & 0  & 0  && 0  & 0  & 0  & 0  && 0  & 0  & 0  & 0  && 0 & 0  & 0  & 0  && 0 & 0  & 0  & 0 \\
&& ZT  && 8 & 25 & 15 & 82 && 5  & 21 & 30 & 82 && 8  & 24 & 22 & 82 && 6 & 25 & 18 & 82 && 6 & 34 & 40 & 82 \\

&&   &&    &    &   &   &&    &    &   &    &&    &    &   &    &&    &    &   &  &&    &    &    &                        \\

Stock market index 
&& DY  && 49 & 59 & 73 & 84 && 43 & 34 & 76 & 84 && 40 & 54 & 73 & 84 && 43 & 43 & 72 & 84 && 21 & 48 & 76 & 84 \\
&& ES  && 71 & 70 & 82 & 84 && 81 & 80 & 84 & 84 && 80 & 75 & 84 & 84 && 77 & 74 & 84 & 84 && 57 & 53 & 84 & 84 \\
&& EW  && 23 & 42 & 62 & 84 && 62 & 65 & 80 & 84 && 38 & 44 & 66 & 84 && 32 & 56 & 60 & 84 && 22 & 42 & 75 & 84 \\
&& FX  && 30 & 44 & 58 & 84 && 38 & 40 & 70 & 84 && 34 & 39 & 60 & 84 && 40 & 41 & 65 & 84 && 21 & 36 & 75 & 84 \\
&& MME && 56 & 57 & 71 & 83 && 56 & 48 & 76 & 83 && 62 & 66 & 77 & 83 && 63 & 62 & 75 & 83 && 49 & 48 & 80 & 83 \\
&& MX  && 42 & 44 & 60 & 84 && 36 & 53 & 76 & 84 && 40 & 40 & 67 & 84 && 49 & 46 & 67 & 84 && 32 & 44 & 79 & 84 \\
&& NQ  && 84 & 84 & 84 & 84 && 41 & 52 & 75 & 84 && 84 & 84 & 84 & 84 && 84 & 84 & 84 & 84 && 10 & 53 & 68 & 84 \\
&& RTY && 47 & 50 & 74 & 84 && 48 & 44 & 72 & 84 && 61 & 56 & 74 & 84 && 44 & 58 & 67 & 84 && 25 & 45 & 70 & 84 \\
&& VX  && 57 & 60 & 67 & 84 && 59 & 56 & 76 & 84 && 61 & 61 & 64 & 84 && 65 & 60 & 71 & 84 && 38 & 43 & 74 & 84 \\
&& X   && 39 & 43 & 58 & 84 && 60 & 55 & 74 & 83 && 33 & 45 & 61 & 84 && 47 & 44 & 68 & 84 && 54 & 63 & 82 & 84 \\
&& XAE && 11 & 31 & 20 & 84 && 12 & 29 & 47 & 84 && 10 & 39 & 33 & 84 && 17 & 42 & 29 & 84 && 76 & 80 & 81 & 84 \\
&& XAF && 11 & 35 & 20 & 84 && 50 & 65 & 66 & 84 && 12 & 31 & 26 & 84 && 18 & 44 & 31 & 84 && 12 & 37 & 44 & 84 \\
&& XAI && 5  & 37 & 19 & 84 && 18 & 44 & 46 & 84 && 8  & 45 & 27 & 84 && 11 & 42 & 29 & 84 && 11 & 38 & 49 & 84 \\
&& YM  && 65 & 75 & 78 & 84 && 83 & 82 & 84 & 84 && 49 & 52 & 79 & 84 && 45 & 54 & 76 & 84 && 72 & 76 & 84 & 84 \\

&&   &&    &    &    &   &&    &    &   &     &&    &    &   &    &&    &    &   &    &&    &    &     &              \\

Six factors 
&& HML && 28 & 45 & 28 & 84 && 55 & 68 & 63 & 84 && 51 & 67 & 50 & 84 && 45 & 62 & 43 & 84 && 61 & 64 & 79 & 84 \\
&& SMB && 17 & 42 & 19 & 84 && 73 & 81 & 68 & 84 && 19 & 40 & 28 & 84 && 13 & 40 & 23 & 84 && 66 & 77 & 64 & 84 \\
&& RMW && 6  & 37 & 14 & 84 && 39 & 58 & 52 & 84 && 5  & 36 & 23 & 84 && 8  & 35 & 20 & 84 && 66 & 73 & 72 & 84 \\
&& CMA && 21 & 40 & 28 & 84 && 15 & 44 & 35 & 84 && 42 & 56 & 45 & 84 && 38 & 56 & 39 & 84 && 63 & 68 & 70 & 84 \\
&& MOM && 29 & 54 & 35 & 84 && 42 & 60 & 56 & 84 && 41 & 61 & 44 & 84 && 38 & 53 & 37 & 84 && 62 & 77 & 68 & 84 \\
&& MKT && 30 & 60 & 67 & 84 && 84 & 84 & 84 & 84 && 40 & 52 & 73 & 84 && 46 & 58 & 70 & 84 && 84 & 84 & 84 & 84 \\
\hline
\end{tabular}
\caption{The number of non-zero monthly integrated beta estimates across 60 factors and five assets over 84 months for the RED-LASSO (RED), ED-LASSO (ED), LASSO, and SVR estimators.}
\label{Table4}
}
\end{table}

\clearpage
\bibliography{myReferences}

\end{spacing}
\end{document}